\documentclass[journal,twoside,web]{ieeecolor}
\usepackage{tmi}
\usepackage{cite}
\usepackage{amsmath,amssymb,amsfonts,dsfont}
\usepackage{algorithmic}
\usepackage{graphicx}
\usepackage{textcomp}
\usepackage{breakcites}
\usepackage{multirow}
\usepackage[colorlinks=true,citecolor=blue,urlcolor=black]{hyperref}
\usepackage{tikz}
\usepackage{textgreek}
\usepackage{booktabs}
\usepackage{tabularx}
\usepackage{xcolor}
\usepackage[misc]{ifsym}

\def\BibTeX{{\rm B\kern-.05em{\sc i\kern-.025em b}\kern-.08em
    T\kern-.1667em\lower.7ex\hbox{E}\kern-.125emX}}
\begin{document}
\title{Topology Optimization in Medical Image Segmentation with Fast $\chi$ Euler Characteristic}
%Topology-aware Segmentation with Euler Characteristic }
\author{Liu Li, Qiang Ma, 
Cheng Ouyang, Johannes C. Paetzold, 
Daniel Rueckert, \IEEEmembership{Fellow, IEEE}, \\Bernhard Kainz, \IEEEmembership{Senior Member, IEEE}
\thanks{\scriptsize L. Li was supported by a Lee Family Scholarship. Q. Ma was supported by a President’s PhD Scholarship from Imperial College London.  Support was also received by the ERC projects Medical Image Analysis with Normative Machine Learning (MIA-NORMAL) 101083647, Deep Learning for Medical Imaging: Learning Clinically Useful Information from Images (Deep4MI) 884622, and The Developing Human Connectome Project (dHCP) data collection 319456. B.K. acknowledges the high-tech agenda Bavaria and HPC resources provided by  the Erlangen National High Performance Computing Center (NHR@FAU, b143dc and b180dc). NHR funding is provided by federal and Bavarian state authorities and partially funded by the German Research Foundation
 (DFG) - 440719683. \\
L. Li, Q. Ma, C. Ouyang, J. Paetzold, D. Rueckert, and B. Kainz are with the Department of Computing, Imperial College London, UK (e-mail: liu.li20$@$imperial.ac.uk).
C. Ouyang is also with Department of Engineering Science, University of Oxford, UK and was with Institute of Clinical Sciences, Imperial College London.
J. Paetzold is also with Weill Cornell Medicine, New York, USA, and Cornell Tech, New York, USA. 
D. Rueckert is the Chair for AI in Medicine and Healthcare, TUM University Hospital, Technical University of Munich, Germany.
B. Kainz is also with SBEIS, King's College London, UK, and the Department AIBE of Friedrich-Alexander-Universität Erlangen-Nürnberg (FAU), Germany.
}}

\maketitle

\begin{abstract}
Deep learning-based medical image segmentation techniques have shown promising results when evaluated based on conventional metrics such as the Dice score or Intersection-over-Union. However, these fully automatic methods often fail to meet clinically acceptable accuracy, especially when topological constraints should be observed, \emph{e.g.}, continuous boundaries or closed surfaces. In medical image segmentation, the correctness of a segmentation in terms of the required topological genus sometimes is even more important than the pixel-wise accuracy. 
Existing topology-aware approaches commonly estimate and constrain the topological structure via the concept of persistent homology (PH). However, these methods are difficult to implement for high dimensional data due to their polynomial computational complexity. To overcome this problem, we propose a novel and fast approach for topology-aware segmentation based on the Euler Characteristic ($\chi$). 
First, we propose a fast formulation for $\chi$ computation in both 2D and 3D. 
The scalar $\chi$ error between the prediction and ground-truth serves as the topological evaluation metric. 
Then we estimate the spatial topology correctness of any segmentation network via a so-called topological violation map, \emph{i.e.}, a detailed map that highlights regions with $\chi$ errors.
% between the prediction and ground-truth. 
Finally, the segmentation results from the arbitrary network are refined based on the topological violation maps by a topology-aware correction network.
Our experiments are conducted on both 2D and 3D datasets and show that our method can significantly 
improve topological correctness while preserving pixel-wise segmentation accuracy.
\end{abstract}


\section{Introduction}
\label{sec:introduction}

\begin{figure}
% \vspace{-0.5cm}
\centering
    \centering
    \includegraphics[width=0.99\linewidth]{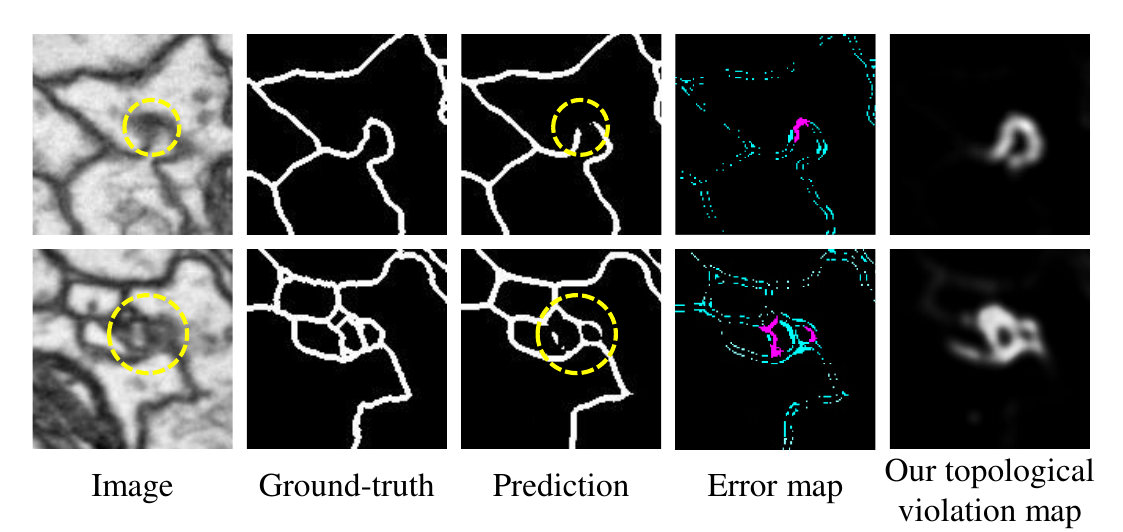}
    \caption{From left to right: image from the CREMI dataset~\cite{funke2018large}, ground-truth map for the neuron boundary, nnU-Net~\cite{isensee2021nnu} segmentation, segmentation error map and our topological violation map. 
    Despite the high Dice accuracy of 88.97\% and 88.85\% for these two examples, the topological errors are very different, \emph{e.g.}, the ring structure is broken in the upper example. In the segmentation error maps, We find that some errors will change the topological structure (highlight in purple) while some errors will not (highlight in blue). Most of these topology-related errors are essentially caused by imaging artifacts, \emph{e.g.}, blurry boundaries, or dark patches. Based on these observations, we design a topology-aware segmentation method that can automatically detect these topology-related errors based on the Euler Characteristic $\chi$. Resulting  topological violation maps are shown in the last column.
    % Left: two segmentations with same Dice score but different topology. red regions highlight the false positive and false negative regions. Middle: vessel segmentation and its surface reconstruction results from nnU-Net. In this example, Dice=0.9, betti error $\beta=(1, 2, 0)$ with $\beta_{gt}=(1, 2, 0)$ and $\beta_{pred}=(x, x, x)$. Right: MR images with the partial volumn effects from dHCP dataset \cite{}, and electron microscopy Images with interpolation blurs in CREMI dataset \cite{}.
    }
\label{fig:fig1}
% \vspace{-0.5cm}
\end{figure}

Performance of automated medical image segmentation methods is commonly evaluated by pixel-wise metrics, \emph{e.g.}, the Dice score or the Intersection-over-Union~\cite{maier2022metrics}. Taking the average accuracy of each pixel/voxel into account, these evaluation methods are well suited to describe the overall segmentation performance. 
However, errors in some regions may be more important than others, \emph{e.g.}, certain errors may lead to the wrong topology of the segemented object, highlighted in purple in Fig.~\ref{fig:fig1} in the $4^{th}$ column. 
In some cases, topological correctness is more important than pixel-wise classification correctness, for example, when segmentation is conducted as a fundamental step for surface reconstruction~\cite{dale1999cortical}. Although the state-of-the-art segmentation methods such as nnU-Net~\cite{isensee2021nnu} and fine-tuned Segment Anything Models (SAM)~\cite{ma2024segment} have proven to achieve high pixel-wise accuracy in many tasks, \emph{e.g.}, brain tumor segmentation~\cite{isensee2021nnu}, prostate analysis~\cite{wang2024direct}, and breast evaluation~\cite{huo2021segmentation}, the topological correctness of the segmentation are usually not addressed.%, for example, the segmentation predictions of neuron boundary are disconnected as shown in Fig.~\ref{fig:fig1}. %Therefore, we propose a topology-aware segmentation method that can improve the topological correctness while preserving pixel-wise accuracy.

%, a misinterpreted topology can cause disconnected surface, and further mislead the downstream clinical analysis.
% However, errors in some regions may be more important than others, \emph{e.g.}, certain errors may lead to misrepresented topology. 
% \textcolor{black}{mislead the downstream clinical analysis}, \emph{e.g.}, \textcolor{black}{in the assessment of the fetal brain development, a misinterpret of the brain surface topology can be caused by only a few incorrect pixels \cite{dale1999cortical}. This misinterpreted topology may mislead clinicians to make  erroneous high-stake decisions}. %certain errors may lead to misrepresented topology. In some cases, topological correctness is more important than pixel-wise classification correctness, such as when segmentation is a fundamental step for surface reconstruction \cite{dale1999cortical}.

% \iffalse
% \fi

As illustrated in Fig.~\ref{fig:fig1}, this paper is motivated by the understanding that only a part of real-world segmentation errors are related to topology. Accordingly, we propose a novel approach where a segmentation network learns to focus more on areas likely to contain topological errors, rather than uniformly addressing all pixel-wise errors equally in an image. % jointly in a topology-aware loss term. 
%in pixel-by-pixel predicted probabilities,.
%Following the motivation that only part of the segmentation errors are topology-related, the general idea of this paper is that instead of evenly optimizing every single error pixel-by-pixel only according to their predicted probability, the network is encouraged to pay more attention to the potential topological error regions.
%Existing topology-aware segmentation methods try to detect topologically critical points utilizing the method of persistent homology (PH).

Current topology-aware segmentation techniques employ persistent homology (PH) to identify critical points that are topologically significant.
%These methods model the segmentation probability map by cubical complex, and PH tracks the topology change of these cubical complex under different filtrations
These techniques represent the segmentation probability map using a cubical complex. PH monitors the changes in topology across various filtrations~\cite{hu2019topology,clough2020topological,clough2019explicit,stucki2023topologically}. Topological  correctness is encouraged by manipulating the probability of these critical points through an additional loss term.

Although integration of these methods is straight forward due to their differential properties, the main challenge arises from the polynomial computational complexity of PH, which limits their applicability to high-dimensional data such as 3D medical images. 
%Despite the innovative design of such differential optimization objective, the main limitation for these methods is the polynomial computation complexity of PH, hindering the application for high dimensional data such as 3D medical images. 
To address this challenge, we propose a fast method based on the Euler characteristic ($\chi$) to detect topologically critical regions that likely contain topological errors.% and further design a topology-aware feature synthesis network to refine the segmentations in these regions.
% that likely contain topological violations. %A topology-aware feature synthesis network is further designed to refine segmentations in these regions.
 % topological violation detection (TVD) block (blue), for training a topology-aware feature synthesis (TFS) network

% For a binary segmentation, $\chi$ is an integer defined as the number of connected components ($\beta_0$) minus the number of holes ($\beta_1$) plus the number of voids in a 3D volume. 
Specifically, we first develop a DL-compatible pipeline for the bit-quads-based 2D $\chi$ calculation~\cite{gray1971local,yao2015novel,arce2022learning}, and further extend this 2D formulation to 3D by introducing bit-octets in Section~\ref{sec:EC3D}. 
% so that our method can be integrated as a guidance function into existing DL pipelines \cite{isensee2018nnu}.
%, using several convolution and pooling layers which are differentiable. 
To identify topologically critical regions, we first calculate $\chi$ for both prediction and ground-truth (GT) locally to get two $\chi$-maps.
Analysing the $L_1$ difference between prediction and GT $\chi$-maps, which we denote as \textit{$\chi$ error map}, we can further reveal more refined details about  topological violations by back-propagating the $\chi$ error with respect to the segmentation, resulting in a \textit{topological violation map} that highlights  topologically critical regions $U$. %Note that the topological violation map maintains the same resolution as the segmentations.

Given this topological violation map, we propose an additional topology-aware feature synthesis (TFS) network to refine segmentations. 
%Considering that a standard segmentation network learns a conditional probability $p(Y|X)$ of segmentation $Y$ based on the input image $X$, 
The motivation for this design emerges from the observation that the topological error regions $\mathbf{U}$ for  segmentation $\mathbf{Y}$ are correlated to imaging artifacts in the input images $\mathbf{I}$, as illustrated in Fig.~\ref{fig:fig1}. Encouraging a segmentation network to directly learn from $\mathbf{U}$ through an additional loss term would be futile since the underlying image features ${I_{ij}, (i,j)\in \mathbf{U}}$ are erroneous. 

Therefore, instead of introducing additional loss terms, our approach relies on a generative TFS network to directly learn a topological prior, $p(\mathbf{Y})$, from regions with topologically correct representations, denoted as $Y_{ij}, (i,j)\in \mathbf{U}^{c}$, where $\mathbf{U}^{c}\cap \mathbf{U}=\varnothing$.

% \iffalse
% The performance of medical image segmentation is closely tied to the precision of the imaging process~\cite{hu2021topology,maier2022metrics}. We observe that there is a correlation between the locations exhibiting topological errors in the segmentation $U$, and the locations where imaging artifacts are present in the input images $V$, as illustrated in Fig.~\ref{fig:fig1}.
% Considering that a standard segmentation network learns a conditional probability $p(Y|X)$ of segmentation $Y$ based on the input image $X$, we question the segmentation network's ability to predict topologically correct values $Y_{ij}$ from an input with imaging artefacts $X_{ij},(i,j)\in V$. 
% Existing topology preserving segmentation methods attempt to encourage a segmentation network to focus more on region $U$ by introducing additional loss terms. However, segmentation networks usually struggle to extract meaningful feature representations from imaging artefacts ${X_{ij}, (i,j)\in V}$, especially when $U$ and $V$ overlap significantly.
% To address this challenge, we propose a topology-aware feature synthesis network to refine the segmentations. This approach directly learns a topological prior, $p(y)$, from regions with topologically correct representations, denoted as $y_{ij}, (i,j)\in U^{c}$, where $U^{c}\cap U=\varnothing$.
% \fi

The primary contributions of this paper are twofold. First, building upon the concept of DL-compatible $\chi$ computation as a tool to inform the network about topologically invariant properties~\cite{li2023robust}, we discuss its applicability in high-dimensional medical data, such as 3D CT and MRI volumes. We define and categorize bit-octets into different groups based on their properties, and explore their combinations to enable fast 3D $\chi$ computation. Second, we conduct experiments on five datasets featuring diverse foreground structures, demonstrating the enhanced segmentation performance of our approach in terms of topological correctness. 

This paper extends some of our initial ideas~\cite{li2023robust}. Firstly, we conduct a comprehensive review of related literature that incorporates priors into segmentation networks, categorizing these priors as either implicit shape priors or explicit topological constraints. Secondly, we broaden the scope of bit-quads-based Euler Characteristic ($\chi$) computation, transitioning from 2D to 3D, by investigating the integration of bit-octets. Thirdly, we evaluate on five different datasets, covering both 2D and 3D domains. 
Note that our approach can be applied to any existing segmentation pipeline as a plug-and-play module. Our code is publicly available\footnote{\scriptsize \href{https://github.com/smilell/Topology-aware-Segmentation-using-Euler-Characteristic}{https://github.com/smilell/Topology-aware-Segmentation-using-Euler-Characteristic}}.

\section{Related work.}
\subsection{Implicit shape priors}
\iffalse
Segmentation networks such as the U-Net~\cite{ronneberger2015u} are typically trained using binary cross-entropy or soft Dice loss functions. %Commonly used loss functions for training such networks is the binary cross-entropy or variants of the Dice score. 
These metrics measure the pixel-wise overlap between the prediction and the GT. 
However, medical image segmentation sometimes trades the integrity of the segmentation prediction for pixel wise accuracy, for example for downstream surface reconstruction tasks.

they do not guarantee the topological correctness. To mitigate this issue, shape priors have
Shape priors have been explored to mitigate this issue, \emph{e.g.}, utilizing shape templates~\cite{mcinerney1996deformable,lee2019tetris} together with diffeomorphic deformations. 
The shape prior are encouraged by constraining the deformation field to be smooth, which relies on the similarity of shape prior with the target.
\fi

% % \cite{lux2024topograph}graph和segmentation结合，

Medical image segmentation aims to provide pixel-level delineation of anatomical structures, facilitating clinical interpretation and downstream analysis. U-Net~\cite{ronneberger2015u} and its variants, such as Swin-Unet~\cite{cao2022swin} and nnUNet~\cite{isensee2021nnu}, have achieved strong performance across various medical imaging tasks. More recently, inspired by the Segment Anything Model (SAM)~\cite{kirillov2023segment} with self-attention mechanisms, MedSAM~\cite{ma2024segment}, MedSAM 2~\cite{zhu2024medical}, and ESP-MedSAM~\cite{xu2024esp} have been developed for medical images, leveraging prompt-based strategies to refine segmentation predictions interactively.

Segmentation networks, however, are typically trained using binary cross-entropy or soft Dice loss functions. These metrics primarily assess pixel-wise overlap between the predicted and ground truth (GT) segmentations. However, in medical image segmentation, prioritizing pixel-wise accuracy can sometimes compromise the structural integrity of predictions, particularly in tasks such as downstream surface reconstruction.

While these methods are effective in certain aspects, they do not explicitly model topological constraints. To address this, shape priors have been explored, for example, by utilizing shape templates~\cite{mcinerney1996deformable, lee2019tetris} with diffeomorphic deformation. These shape priors are reinforced by constraining the deformation field to be smooth, which assumes a similarity between the shape prior and the target.
%If an initially topologically correct shape template is diffeomorphically deformed to match a target segmentation, through, \emph{e.g.}, a deep neural network parameterisation~\cite{lee2019tetris}, then the resulting deformed shape is guaranteed to be topologically correct.  
However, a limitation of these methods is their reliance on predictions being extremely close to the shape priors, a condition that is often unattainable in practical scenarios. 

Other approaches explicitly incorporate shape information into the training process~\cite{oktay2017anatomically}.  Additionally, implicit shape awareness, based on boundary and centerline information that can model connectivity, has been explored in recent studies~\cite{kervadec2019boundary,shit2021cldice,kirchhoff2024skeleton}. Despite these advancements, these methods do not inherently derive from topological principles and, as such, do not explicitly model topological information.

\subsection{Explicit topological constraints}

Topology in segmentation has been explored through the use of Persistent Homology (PH) \cite{hu2019topology,clough2020topological,clough2019explicit,li2022fetal,de2021segmentation,byrne2022persistent,stucki2023topologically,berger2024topologically}. PH identifies topological features by tracking the emergence and dissolution of connected components, holes, and voids as filtration changes. In practice, PH-based methods detect critical points associated with topological changes and manipulate their values to align the predicted topology with the ground truth (GT). This approach can be made differentiable by modelling the probability of critical points in the segmentation. 
Recently, PH loss has also been applied to multi-category segmentation where it models the combinations of different labels to enforce topological constraints~\cite{santhirasekaram2023topology,berger2024topologically}. Beyond segmentation tasks, PH has also been applied in image generation to guide the synthesis of images with topologically correct structures~\cite{abousamra2023topology,xu2024topocellgen}.
% \textcolor{black}{}

Despite optimised computational packages for PH \cite{kaji2020cubical,maria2014gudhi}, its polynomial computational complexity remains a major limitation, especially for large-scale or higher dimensional datasets \cite{hu2022structure}. Further, PH-based loss optimization in each iteration targets only a limited number of critical points. Newer approximation methods, such as those based on Discrete Morse Theory \cite{hu2021topology} and Homotopy Warping \cite{hu2022structure}, have reduced computational complexity to $O(n\log n)$ and $O(n)$, respectively.

Nevertheless, these approaches primarily focus the network's attention on regions with topological discrepancies when compared to the GT, still relying on a pixel-wise topology loss term instead of learning a global topological representation. Moreover, in medical image segmentation, topological errors often result from imaging artifacts, such as blurred boundaries. Training the network on these compromised inputs may not yield improved performance. To address this gap, we introduce a $\chi$-based method that directly learns from the label map space, avoiding the lack of image space representation in those imaging artifact regions.% Leveraging the integration of deep learning techniques in $\chi$ computation, our method is capable of achieving real-time performance.

% Persistent homology (PH) is a technique that extends EC to study the evolution of topological space. The evolution is parameterized by a filtration $f: \mathcal K \rightarrow \mathbb{R}$, which captures how the space changes. 
% PH extracts topological features by tracking the birth and death of connected components, holes, and voids as the filtration parameter varies. Although PH is a powerful tool for capturing the topological features of a space, its polynomial computational complexity is a limitation that hinders its use in large-scale datasets.
\begin{figure}
\centering
\includegraphics[width=0.7\linewidth]{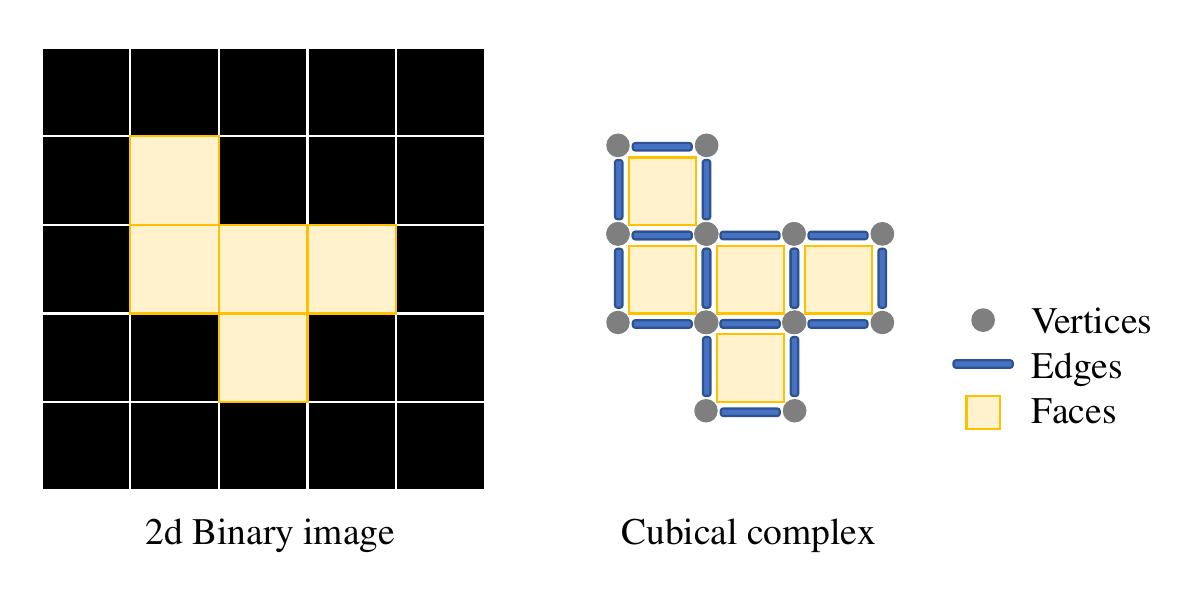}
\caption{Construction of a cubical complex from a 2D binary image. Yellow and black blocks represent foreground and background segmentations, respectively.} 
\label{fig:cubical}
\end{figure} 

\begin{figure}
% \vspace{-0.5cm}
\centering
    \centering
    \includegraphics[width=0.99\linewidth]{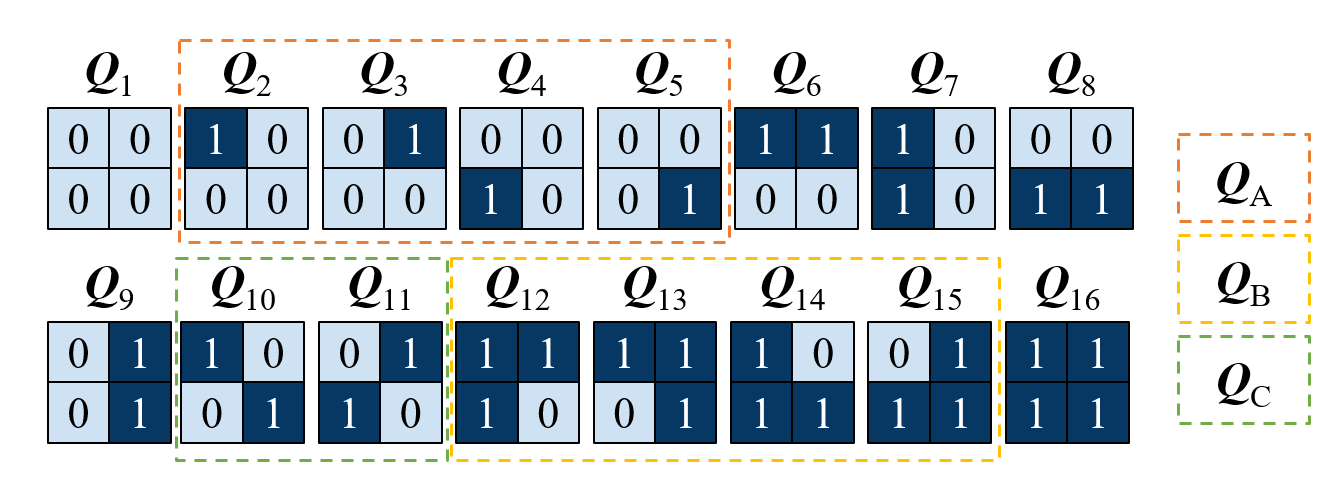}
    \caption{Bit-quads for 2D $\chi$ computation.}
\label{fig:bitquad}
% \vspace{-0.5cm}
\end{figure}

\begin{figure*}
% \vspace{-0.5cm}
\centering
    \centering
    \includegraphics[width=0.9\linewidth]{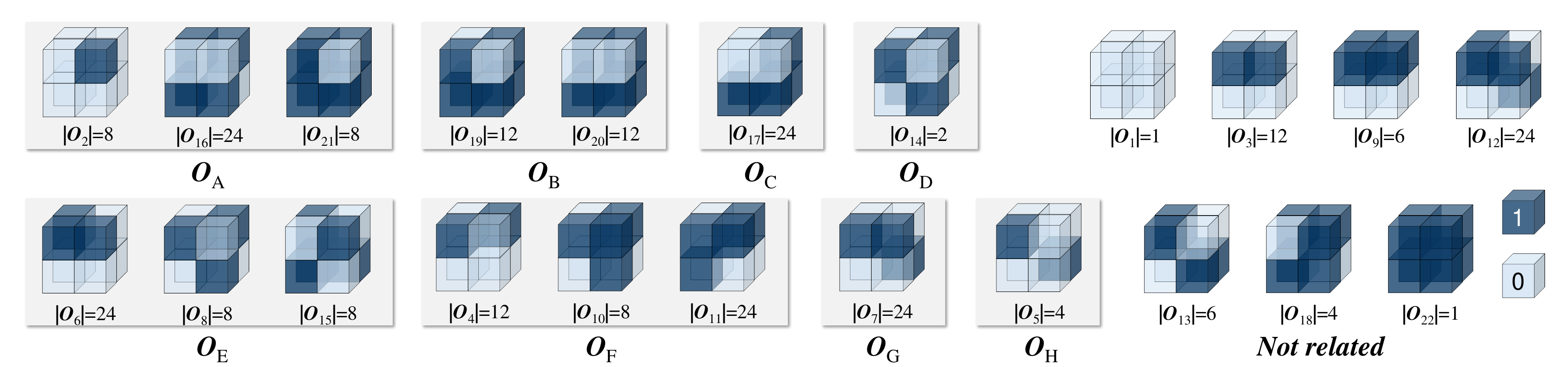}
    \caption{Bit-octets for 3D $\chi$ computation. $\vert O \vert$ represents the cardinality of the set $O$.}
\label{fig:Bit-octets}
% \vspace{-0.5cm}
\end{figure*}

\begin{figure*}
% \vspace{-0.5cm}
\centering
    \centering
    \includegraphics[width=0.95\linewidth]{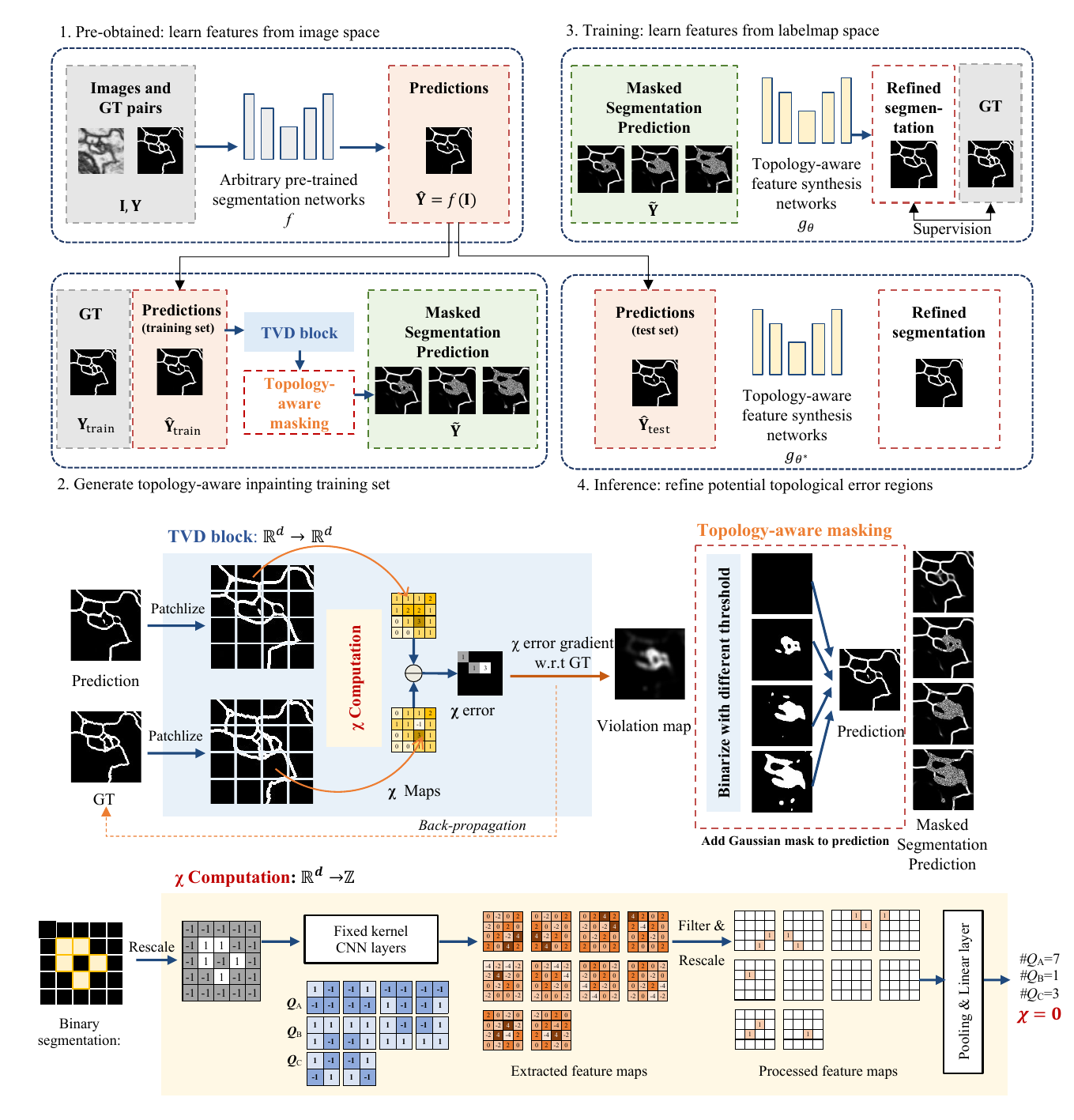}
    \caption{Overview of our method. Given arbitrary pre-trained segmentation networks $f$ and its predictions $\widehat{\mathbf{Y}}$, we generate the topology-aware masked segmentation predictions $\widetilde{\mathbf{Y}}$ (green) from the topological violation detection (TVD) block (blue), for training a topology-aware feature synthesis (TFS) network $g_\theta$. During inference, the raw predictions from the test set $\widehat{\mathbf{Y}}_{\rm{test}}$ are corrected by the trained TFS network $g_\theta^{*}$. In TVD block, the $\chi$-maps are computed as shown in the yellow region.
    }

\label{fig:pipeline}
% \vspace{-0.5cm}
\end{figure*}

\section{Method}

% \textit{Notations}
% topological space $X$
% dimension $k$ maximum dimension $K$
% Simplicial complex, cubical complex $\mathcal{K}$

% k-dimensional homology groups $H_k(X)$, what is $X$

% betti number $\beta_k\in\mathbb{N}$

% Euler Characteristic $\chi\in\mathbb{Z}$
% the number of $k$-dimensional cells $C_i$           % used to be $N_k$
% bit-quad $Q_i$ categorized to $Q_A$                 % used to be  $\mathbf{K}_1$, $\mathbf{K}_2$ and $\mathbf{K}_3$ 
% the number of occurrence of bit-quads: $\#Q_i$                   % used to be $M_l$, $l\in\{1,2,3\}$
% bit-octets: $O_i$ and $\#O_i$  

% ~\cite{chazal2021introduction,munch2017user,wasserman2018topological,EMApCourse,schenckalgebraic,freedman2009algebraic,artin_algebra_1991}. 

% \noindent\textbf{Preliminaries.}
\subsection{Preliminaries}
Algebraic topology studies the properties of topological spaces~\cite{wasserman2018topological,eilenberg2015foundations}. This field aims to identify and characterize topological invariants, such as the number of holes, the connectivity, or the homotopy groups of a space~\cite{chazal2021introduction,munch2017user,freedman2009algebraic}. 
In algebraic topology, high-dimensional data, such as point clouds, can be modeled using a simplicial complex $\mathcal{K}$, which is a set composed of points, line segments, triangles, tetrahedra, and higher-dimensional equivalents. 

Image segmentation for 2D or 3D gray-scale images is defined on grids $\mathbf{I}\in \mathbb{R}^{h\times w \times l}$. This can be naturally modeled by cubical complexes $\mathcal{K}$ to reduce the complexity of arbitrary possible connectivity between different vertices~\cite{freedman2009algebraic}. The basic elements of a $3$-dimensional cubical complex $\mathcal{K}$ are points, line segments, squares, and cubes. 
For example, in a 3D binary image, each voxel is modeled by a cube, and every face of this cube, \emph{i.e.}, 8 points, 8 line segments, and 4 squares, are also in this cubical complex $\mathcal{K}$, as shown in Fig.~\ref{fig:cubical}.

\noindent\textbf{Betti Numbers.}
Betti numbers $\beta_k$ are topological invariants that describe the property of geometric objects that remain invariant under homeomorphisms, \emph{i.e.},  transformations include stretching, bending, and deforming without tearing or gluing~\cite{wasserman2018topological}. Specifically, Betti numbers describe how many ``holes'' exist in a topology space $X$ in dimension $k$. 
Mathematically, Betti numbers are defined as the ranks of the $k$-dimensional homology groups $H_k(X)$. Specifically, the $k$-dimensional homology group is a collection of $k$-dimensional non-boundary cycles, which can be intuitively understood as ``holes''~\cite{freedman2009algebraic}.
%All these non-boundary cycles can be categorized into different homology classes according to the standard that if they can be freely deformed into each other without breaking their topological structure.
All non-boundary cycles can be grouped into homology classes based on whether they can be freely transformed into one another without altering their topological structure. 
The rank of $H_k(X)$ captures how many independent ``holes'' exist in this homology group. In dimension 0, 1 and 2, Betti numbers $\beta_k$ represent the number of connected components, holes, and voids, respectively.
Any $k$-dimensional cubical complex can only have $(k-1)$-dimensional non-zero Betti numbers, \emph{i.e.}, $(k-1)$-dimensional holes.
Betti numbers are also commonly used to evaluate the topological correctness. The Betti error $e$ is defined by the averaged $L_1$ distance between two Betti numbers in all dimension $k$ as 
\begin{equation}
e_k = \|\beta_k^1 - \beta_k^2 \|_1 \text{ and }  e = \frac{1}{K}\sum_{k=0}^{K-1} e_k.
\end{equation}
% group: set and operation, identity (operation) element = element
% generator of group: is a minimal subset of group, in Z 1 (+)
% rank of group: number of elements in the generator (a subset)

% homology group: free group infinit elements
% boundary operator: k-dimension k-simplex -> (k-1)-simplex: apply boundary twice vanish

% kernel of boundary operator: whose boundary is zero, all closed loop
% image of boundary operator: who can be a boundary

% ; and $\beta_0$, $\beta_1$ and $\beta_2$ have the meaning of the number of connected components, holes and voids. Note that for any 3D images defined on a 3-dimensional space, $\beta_k(\mathcal{K})=0$ when $k\ge3$.

% For image segmentation, gray-scale images defined on grids $\mathbf{I}\in \mathbb{R}^{h\times w}$ are easier to analyze with a cubical complex instead of simplicial complex. The basic elements of a cubical complex are points, line segments, squares, and cubes. Fig.~1 in the appendix shows how a gray-scale image $\mathbf{I}\in \mathbb{R}^{h\times w}$ can be modeled using a cubical complex $\mathcal{K}$.

\noindent\textbf{Euler Characteristic.}  $\chi\in\mathbb{Z}$ is a topological invariant that can be used to distinguish between different shapes of objects; it is commonly used to help understand and describe their structure~\cite{richardson2014efficient}.
% that describes the structure and properties of a given topological space.
% For a %simplicial/ 
% cubical complex $\mathcal{K}$ defined in a $k$-dimensional space, EC $\chi$ is 
$\chi$ is defined as the alternating sum of the number of $k$-dimensional components $N_k$, \emph{e.g.}, the number of vertices, edges, faces, etc.
$\chi$ also equals to the alternating sum of Betti numbers $\beta_k$:
%. Mathematically, EC can be formalized as:
% of k-dimensional Betti numbers $\beta_k$, which are the ranks of the homology groups: 
\begin{equation}
    \chi(\mathcal{K})=\sum_{k=0}^n (-1)^k N_k = \sum_{k=0}^{n-1} (-1)^k \beta_k(\mathcal{K}).
    \label{eq:0}
\end{equation}

% , or alternating sum of the ranks of the k-dimensional homology groups, called Betti number $\beta_k$

% $\chi(\mathcal{K})=\sum_{k=0}^n (-1)^k N_k$, where $N_0$, $N_1$ and $N_2$ are the number of vertices, edges and faces, respectively. 
% Alternatively, EC can be expressed in terms of Betti numbers $\beta_k$, which are the ranks of the homology groups of the space: $\chi(\mathcal{K})=\sum_{k=0}^n (-1)^k \beta_k(\mathcal{K})$. 
Specifically, for $3$-dimensional images $\chi(\mathcal{K}) = N_0 - N_1 + N_2 - N_3= \beta_0- \beta_1 + \beta_2$, where $N_0$, $N_1$, $N_2$ and $N_3$ are the number of vertices, edges, faces and squares.

% xxx compare EC with betti number and PH
% Betti numbers and its related PH is a common way to express the the topology of a space, where 

% $Q_i$ categorized to $Q_A$    

% \iffalse

% \fi
    % in three steps: (1) We first generate an initial prediction from the segmentation network, where topological errors may occur. (2) Second, we propose a TVD block to detect topology-violating regions. (3) Based on this violation map, we use a topology-aware feature synthesis network to correct errors and improve the segmentation.}

\noindent\textbf{Bit-quad-based method for $\chi$ computation.} Conventional approaches such as the Gray algorithm have been developed to calculate $\chi$ in 2D binary images~\cite{gray1971local,yao2015novel}. Instead of directly counting the number of vertices $N_0$, edges $N_1$, and faces $N_2$, the Gray algorithm proves that $\chi$ in 2D binary images equals to a linear combination of the occurrences of different $2\times 2$ patterns $Q_i, i\in\{1,2,..., 2^4\}$, called bit-quads, as shown in Fig.~\ref{fig:bitquad}. 
Note that the combination of bit-quads can have more than one solution and not all the $2^4=16$ bit-quads will contribute to $\chi$. 
% however from the literature review, the EC can be calculated only by the occurrences of a few bit-quads. 
For example in the Gray method, $\chi$ can be computed from the occurrences of only 10 bit-quads $\#Q_i$ as a linear combination $f$:
\begin{equation}
    \chi = f(\#Q_i) = \frac{1}{4}(\#Q_\mathrm{A} - \#Q_\mathrm{B} - 2\#Q_\mathrm{C}),
    \label{eq:1}
\end{equation}

where 
\begin{align}
Q_\mathrm{A} &=\{Q_2, Q_3, Q_4, Q_5\}\\
Q_\mathrm{B} &=\{Q_{12}, Q_{13}, Q_{14}, Q_{15}\}\\
Q_\mathrm{C} &=\{Q_{10}, Q_{11}\}.
\end{align}

\subsection{Extension of $\chi$ computation to 3D}\label{sec:EC3D}
Inspired by the Grey method for computing $\chi$ in 2D via a linear combination of the occurrences of bit-quads, we extend this concept to 3D by exploring whether a similar linear combination of the occurrences of 3D bit-octets can represent $\chi$ in 3D. 
We define 3D bit-octets as $O_j$, with their respective occurrences denoted as $\#O_j$.

The problem then reduces to determining whether there exists a coefficient vector $\mathbf{\omega}=[\omega_1, \omega_2, ..., \omega_N]$ such that for any 3D binary segmentation $\mathbf{Y}_i$, the following equation satisfies:
\begin{equation}
    \chi_i = \sum_{j=1}^{N} \#O_{i,j}\mathbf{\omega}_j, %\text{s.t.} \quad N = 256, \quad M \gg N,
    \label{eq:ec3d}
\end{equation}
where  $N$  is the number of bit-octets, % $M$  is the number of equations in the system, 
$\chi_i$ is the Euler characteristic of segmentation $\mathbf{Y}_i$, and $\#O_{i,j}$ is the occurrence of the $j^\mathrm{th}$ bit-octet in segmentation $\mathbf{Y}_i$. 
Binary $2\times 2\times 2$ bit-octets have $N=2^8$ different combinations. However, since $\chi$ is a topological invariant that it remains unchanged under affine transformations including rotation and flipping. 
Consequently, symmetric bit-octets must shares the same coefficients $\omega_j$. 
By categorizing bit-octets, the number of independent $\omega_j$ is reduced to $N=22$, denoted as $\#O = [\#O_1, \#O_2, ... \#O_{22}]$. The representative bit-octets of each group are illustrated in Fig.~\ref{fig:Bit-octets}.

\textbf{Empirical Verification and Optimization.} To verify the existence of a valid coefficient set $\mathbf{\omega}=[\omega_1, \omega_2, ..., \omega_{22}]$, we generate $M$ pairs of $\{\chi_i, \#O_i\}$ from $M$ 3D binary volumes, ensuring  $M \gg N$. We solve the overdetermined linear system to obtain a feasible combination of coefficients that satisfies all M samples:
\begin{equation}
\begin{split}
    \chi = \frac{1}{8}&(\#O_\mathrm{A} + 2\#O_\mathrm{B} + 3\#O_\mathrm{C} +4\#O_\mathrm{D} -\#O_\mathrm{E}\\&- 2\#O_\mathrm{F} - 3\#O_\mathrm{G} - 5\#O_\mathrm{H}),
    \label{eq:chi_3d}
\end{split}
\end{equation}
where $O_\mathrm{A} =\{O_2, O_{16}, O_{21}\}$, $O_\mathrm{B} =\{O_{19}, O_{20}\}$, $O_\mathrm{C} =\{O_{17}\}$, $O_\mathrm{D} =\{O_{14}\}$, $O_\mathrm{E} =\{O_{6}, O_{8}, O_{15}\}$, $O_\mathrm{F} =\{O_{4}, O_{10}, O_{11}\}$, $O_\mathrm{G} =\{O_{7}\}$, and $O_\mathrm{H} =\{O_{5}\}$. 
As a result, we found a linear combination of bit-octets to calculate 3D $\chi$ that satisfies all samples in our dataset. This derived formula enables efficient computation of $\chi$ in 3D binary segmentations.

\subsection{Realization}
We illustrate our method in Fig.~\ref{fig:pipeline}, which consists of three components: (1) a segmentation network, (2) a \textit{topological violation detection} (TVD) block and (3) a \textit{topology-aware feature synthesis} (TFS) network.

In (1), we utilize a U-Net to predict a segmentation probability map $\widehat{\mathbf{Y}}\in[0,1]^{h\times w \times d}$, 
%$\widehat{\mathbf{Y}}\in\mathbb{R}^{h\times w \times d}$
 with the supervision of GT segmentation $\mathbf{Y}\in\{0,1\}^{h\times w \times d}$. 
 The predicted segmentation map may contain violations of a given topological prior.
 %We note that this network can be replaced by any other existing segmentation pipelines. 
(2) is the main contribution of our approach, consisting of two sub-nets: a \textit{$\chi$-Net} and a \textit{Visualization Net}. 
The $\chi$-Net takes the predicted $\widehat{\mathbf{Y}}$ and GT segmentation $\mathbf{Y}$ as inputs and predicts their corresponding $\chi$-maps $\widehat{\mathbf{X}}\in\mathbb{Z}^{h\times w \times d}$ and $\mathbf{X}\in\mathbb{Z}^{h\times w \times d}$. We then measure the $\chi$ error $e\in\mathbb{R}$ as $L_1$ distance with $e=\Vert \mathbf{X} - \widehat{\mathbf{X}}\Vert_{1}$. % between $\widehat{\mathbf{E}}$ and $\mathbf{E}$. 
The Visualization Net takes $e$ as input and produces a topology violation map $\mathbf{V}\in\mathbb{R}^{h\times w \times d}$, which highlights the regions with topological errors. 
Finally, in (3), we design a TFS network, which learns to fill the missing information in the erroneous regions. This subnetwork takes the predicted segmentation $\widehat{\mathbf{Y}}$ and violation map $\mathbf{V}$ as input and generates a topology-preserving segmentation%$\widetilde{\mathbf{Y}}$
, which is the final prediction. 

% During training, we use the TVD block (red arrows in Fig.~\ref{fig:pipeline}) to generate the violation maps $\mathbf{V}$ to further guide the next feature synthesis network. During inference, we only run the first segmentation network and TFS network, as indicated by the upper blue arrows in Fig.~\ref{fig:pipeline} to produce the final topology-preserving segmentation results.

% and further visualize the Euler error between the predicted and ground truth segmentations by computing the gradient of the Euler error with respect to the segmentations.

%\subsection{Topological Violation Detection}

%
%--
\noindent\textbf{Topology-violation detection (TVD).}
TVD consists of two parts: a $\chi$-Net and a Visualization Net (Fig.~\ref{fig:pipeline} bottom). 
The Gray algorithm to calculate $\chi$ cannot be directly integrated into the gradient-based optimization process as is not differentiable. To overcome this problem,  we propose a DL-compatible $\chi$-Net as a CNN-based method that leverages the Gray algorithm to calculate $\chi$. The $\chi$-Net serves as a function that maps the segmentation space to the $\chi$ space $g: \mathbf{Y} \rightarrow \chi$.
It is worth mentioning that in order to preserve spatial information, for each 2D segmentation $\mathbf{Y}$, $\chi$-Net locally produces Euler numbers $\chi_{ij}$ with the input of a segmentation patch $\mathbf{P}_{ij}^{\rm {\Delta}} = \mathbf{Y}_{i:i+ {\rm {\Delta}}, j:j+{\rm {\Delta}}}$, where ${\rm {\Delta}}$ is the patch size. We can therefore obtain a $\chi$-map $\mathbf{X}$ with each of its elements $\chi_{ij}$. Similar process is also applied for 3D segmentation.

% EC-Net generates an Euler map $\mathbf{X}$, where each of its element $\chi_{ij}$ at coordinate $[i, j]$ is a local Euler number corresponding to a segmentation patch $\mathbf{P}_{ij,{\rm {\Delta}}} = \mathbf{Y}_{i:i+ {\rm {\Delta}}, j:j+{\rm {\Delta}}}$, where ${\rm {\Delta}}$ is the patch size.

% euler number is calculated based on segmentation patches $\mathbf{Y}[i:i+\Delta, j:j+\Delta]$, where $\Delta$ is the patch size. Therefore, in our setting,

$\chi$-Net consists of three parts: 1) fixed kernel CNN layers, 2) an average pooling layer and 3) a linear transformation $f$. 
For 2D segmentations, we first utilize three CNN layers with fixed kernels to localize the bit-quads in the segmentation following the Gray algorithm.
The values of the kernels are the same as the bit-quads $\mathbf{K}_{\mathrm{A}}\in \{-1,1\}^{2\times2\times4}$, $\mathbf{K}_{\mathrm{B}}\in \{-1,1\}^{2\times2\times4}$, and $\mathbf{K}_{\mathrm{C}}\in \{-1,1\}^{2\times2\times2}$, as shown in Fig.~\ref{fig:pipeline}. 
Note that we first binarize the prediction probability map $\widehat{\mathbf{Y}}$ and further normalize it to $\{-1, 1\}$. Therefore, if and only if the prediction has the same pattern as the bit-quads, it will be activated to $4$ after convolution. % to make sure that we only count the activated values that equal to 4. 
Subsequently, we apply an average pooling layer to obtain the local number of bit-quads $\#Q_{l, ij}$ at coordinate $(i,j)$. The process can be summarized as 
\begin{equation}
\#Q_{l, ij} = {\rm {\Delta}} ^{2} \cdot {\rm {AvgPool}} ( \mathbb{I}(\mathbf{P}_{ij}^{\rm {\Delta}} * \mathbf{K}_l=4)),
\label{eq:3}
\end{equation}
where $l \in \{\mathrm{A,B,C}\}$ is the index of 2D kernel groups, $*$ represents the convolutional operation, and $\mathbb{I}(\cdot)$ is an indicator function that equals 1 if and only if the input is true. Note that the patch size of average pooling is the same as the patch size of the segmentation~${\rm {\Delta}}$. 
Finally, following Eq.~\ref{eq:1}, a linear transformation is used to calculate the $\chi$-map $\mathbf{X}$ with each of its element as $\chi_{ij} = \frac{1}{4}(\#Q_{\mathrm{A},ij} - \#Q_{\mathrm{B},ij} - 2\#Q_{\mathrm{C},ij})$. During training, we separately take both $\mathbf{Y}$ and $\widehat{\mathbf{Y}}$ as the input of $\chi$-Net and obtain their corresponding $\chi$-maps $\mathbf{X}$ and $\widehat{\mathbf{X}}$. Similar process is also applied to 3D bit-octets.%, with 22 groups of bit-octets.
% The calculations in the EC-Net are all differentiable.

% \begin{equation}
%     \mathbf{X}[ij] = g(\mathbf{Y}[i:i+\Delta, j:j+\Delta]) =f(M_1, M_2, M_3),
%     \label{eq:2}
% \end{equation}
% where $S_{ij}$ represents $\mathbf{Y}[i:i+\Delta, j:j+\Delta]$ as a segmentation patch.

In the second part of TVD, we measure the $\chi$ error as $L_1$ distance with $e=\Vert \mathbf{X} - \widehat{\mathbf{X}}\Vert_{1}$. To obtain a high-resolution indication map that highlights the regions of the input map contributing to the $\chi$ error, 
we calculate the gradient of $e$ with respect to the segmentation maps. This gradient $\mathbf{V}= \partial e/ \partial \mathbf{Y}$, referred to as the topology violation map, provides a detailed visualization of the $\chi$ error and is the output of TVD. % +  

\noindent\textbf{Topology-aware feature synthesis (TFS).}
Here we aim to improve the topological correctness of segmentation by utilizing the detected topology violation map. The motivation is that topological errors are often caused by poor feature representation in the input image, such as blurry boundary regions. These errors are difficult to correct when training directly from the image space. To address this, we propose a TFS network that explicitly learns how to repair the topological structures from the segmentation itself. 

During training, the topology violation map is first binarized by a threshold $t$ to highlight regions with topology violation:
\begin{equation}
    \mathbf{V}^t = \mathbb{I}(\mathbf{V}\geq t),
\end{equation}
where a lower threshold results in more pixels being classified as topologically incorrect. Then, we construct the masked segmentation map $\widetilde{{\mathbf{Y}}}$ (with its element $\tilde{Y}_{i,j}$) by replacing pixels identified in $\mathbf{V}^t$ with Gaussian noise, while keeping the unmasked pixels unchanged. The noise $\epsilon \sim \mathcal{N}(0, 1)$ is normalized to the range $(0, 1)$ using a Sigmoid function:
\begin{equation}
\tilde{Y}_{i,j} = \left\{
\begin{aligned}
\frac{1}{1+ e^{-\epsilon}} &, & V^t_{i,j} = 1,\\
\hat{Y}_{i,j} &, &  V^t_{i,j} = 0.
\end{aligned}
\right.
\label{eq:4}
\end{equation}
The resulting masked segmentation map $\widetilde{{\mathbf{Y}}}$ is then fed into the TFS network as input, which is trained to restore the masked regions under GT supervision with correct topology. 

During inference, we feed the feature synthesis network with real predictions $\widehat{\mathbf{Y}}$. We show evidence for the effectiveness of this design in Section~\ref{sect:eval}.

\section{Evaluation}
\label{sect:eval}

\begin{table*}[t]   %htbp
  \centering
  \caption{%\MakeLowercase{
  Segmentation performance of our method on CREMI, ISBI13, FIVES, dHCP, and TopCoW datasets with two ablation experiments (Ours without TVD and TFS). $e$, $e_0$ and $e_1$ represent Betti error, Betti error 0 and Betti error 1. ASD is average surface distance. Mean value and standard deviation are reported in each term. Statistical significance levels are reported by p-value. Best values are highlighted in \textbf{bold}. 
  % }
  }
  {\scriptsize
    \begin{tabularx}{\textwidth}{p{0.2cm}p{2.1cm}p{2.1cm}cp{2.7cm}p{2.7cm}p{2.5cm}p{2.3cm}}%[WIP no touching]Xc
    \toprule
    Data & Method & Dice $\uparrow$ & &$e$ $\downarrow$& $e_0$ $\downarrow$ & $e_1$  $\downarrow$ &  ASD $\downarrow$  \\ %& & s/batch $\downarrow$ \\ %& Hausdoff $\downarrow$  &
    \midrule
    \multirow{9}[2]{*}{\rotatebox{90}{CREMI (2D)}} 
    %24.51 \\
          & cl-Dice loss~\cite{shit2021cldice}           & $83.48_{\pm 7.31\,  (p < 0.001)} $  & & $10.403_{\pm5.814 \, (p < 0.001)}$ & $7.445_{\pm 5.224 \, (p < 0.001)}$ & $2.958_{\pm 2.975 \,  (p < 0.001)}$ &  $1.612_{\pm 4.448 \,  (p < 0.001)}$  \\ %& &  0.047 \\  & 49.517 &
          & Boundary loss~\cite{kervadec2019boundary}    & $84.00_{\pm 7.20\,  (p < 0.001)}$   & & $9.877_{\pm 5.506 \, (p < 0.001)}$ & $6.938_{\pm 4.840 \, (p < 0.001)}$ & $2.939_{\pm 3.025\,  (p < 0.001)}$ &  $1.612_{\pm 4.351\,  (p < 0.001)}$ \\ %& & \textbf{0.041} \\ & 49.460 & 
          & PH loss~\cite{hu2019topology}                & $84.07_{\pm 7.30\,  (p < 0.001)}$   & & $9.103_{\pm 5.175 \, (p < 0.001)}$ & $6.213_{\pm 4.667 \, (p < 0.001)}$ & $2.889_{\pm 2.986\,  (p < 0.001)}$ &  $1.690_{\pm 4.607\,  (p < 0.001)}$ \\ %& & 8.772 \\ & 49.849 & 
          & Warp loss~\cite{hu2022structure}             & $83.03_{\pm 7.32\,  (p < 0.001)}$   & & $11.781_{\pm 6.572 \, (p < 0.001)}$ & $8.877_{\pm 6.066 \, (p < 0.001)}$ & $2.905_{\pm 2.996\,  (p < 0.001)}$ &   $1.634_{\pm 4.453\,  (p < 0.001)}$ \\ %& & 0.852 \\ & 50.347 &

           & Swin-Unet~\cite{cao2022swin}                & $84.13_{\pm 5.24 \, (p < 0.001)}$   & & $7.127_{\pm 4.770 \, (p < 0.001)}$ & $4.878_{\pm 4.044 \, (p < 0.001)}$ & $2.249_{\pm 2.454\,  (p < 0.001)}$ &   $\mathbf{0.756}_{\pm 2.562\,  (p < 0.001)}$ \\  
           & MedSAM~\cite{ma2024segment}                 & $82.24_{\pm 8.92 \, (p < 0.001)}$& & $8.794_{\pm 25.827 \, (p < 0.001)}$ & $6.483_{\pm 22.048 \, (p < 0.001)}$ & $2.311_{\pm 4.716\,  (p < 0.001)}$ &   $1.934_{\pm 6.045\,  (p < 0.001)}$ \\  
          \cmidrule{2-8}
          & Ours w/o TVD+TFS                             &  $83.92_{\pm 7.11 \, (p < 0.001)}$  & & $10.787_{\pm 5.985 \, (p < 0.001)}$ & $7.926_{\pm 5.528 \, (p < 0.001)}$ &  $2.861_{\pm 2.913\,  (p < 0.001)}$ &   $1.571_{\pm 4.418\,  (p < 0.001)}$ \\ %& & \textbf{0.041} \\ & 50.284 &
          & Ours w/o TVD                                 & $85.19_{\pm 6.65 \, (p < 0.001)}$   & & $5.191_{\pm 3.615 \, (p < 0.001)}$ & $3.009_{\pm 2.860 \, (p < 0.001)}$ &  $2.182_{\pm 2.417\,  (p < 0.001)}$ &   $1.298_{\pm 3.988\,  (p < 0.001)}$ \\%&  & %0.083 \\ & \textbf{40.503} &
          & Ours                                         & $\mathbf{85.24}_{\pm 6.64}$          & & $\mathbf{4.941}_{\pm 3.418}$ & $\mathbf{2.799}_{\pm 2.666}$ & $\mathbf{2.143}_{\pm 2.404}$ &  $1.307_{\pm 4.028}$\\ % &  & 0.189 \\ & 41.301 & 
        \midrule
    \multirow{9}[2]{*}{\rotatebox{90}{ISBI13 (2D)}} 
          & cl-Dice loss~\cite{shit2021cldice}           & $80.66_{\pm 5.71 \,  (p < 0.001)}$   & & $3.343_{\pm 2.359 \,  (p < 0.001)}$ & $1.813_{\pm 1.829 \, (p < 0.001)}$ & $1.530_{\pm 1.372\,  (p < 0.001)}$ &  $0.751_{\pm 0.764\,  (p < 0.001)}$\\ % & &  0 \\ & 37.122 & 
          & Boundary loss~\cite{kervadec2019boundary}    & $80.18_{\pm 6.84 \,  (p < 0.001)}$   & & $4.060_{\pm 2.766 \,  (p < 0.001)}$ & $2.567_{\pm 2.342 \, (p < 0.001)}$ & $1.493_{\pm 1.340\,  (p < 0.001)}$ &   $0.852_{\pm 1.098\,  (p < 0.001)}$ \\ %& &  0 \\& 36.928 & 
          & PH loss~\cite{hu2019topology}                & $80.26_{\pm 6.74 \,  (p < 0.001)}$   & & $3.973_{\pm 2.752 \,  (p < 0.001)}$ & $2.403_{\pm 2.260 \, (p < 0.001)}$ & $1.570_{\pm 1.390\,  (p < 0.001)}$ &   $0.783_{\pm 0.954\,  (p < 0.001)}$ \\ %& &  0 \\& 35.971 & 
          & Warp loss~\cite{hu2022structure}             & $80.32_{\pm 6.69 \,  (p < 0.001)}$   & & $5.420_{\pm 3.408 \,  (p < 0.001)}$ & $3.593_{\pm 3.151 \, (p < 0.001)}$ & $1.827_{\pm 1.656\,  (p < 0.001)}$ &   $0.759_{\pm 0.888\,  (p < 0.001)}$\\ % & &  0 \\& 36.605 & 

        & Swin-Unet~\cite{cao2022swin} & $80.18_{\pm 6.41 \,  (p < 0.001)}$ & & $5.643_{\pm 3.784 \,  (p < 0.001)}$ & $2.797_{\pm 2.410 \, (p < 0.001)}$ & $2.847_{\pm 2.577\,  (p < 0.001)}$ &   $\mathbf{0.514}_{\pm 0.336\,  (p < 0.001)}$ \\  
         & MedSAM~\cite{ma2024segment} & $80.40_{\pm 5.64 \, (p < 0.001)}$ & & $5.190_{\pm 4.084 \, (p < 0.001)}$ & $2.777_{\pm 2.857 \, (p < 0.001)}$ & $2.413_{\pm 2.723\,  (p < 0.001)}$ &   $0.667_{\pm 0.618\,  (p < 0.001)}$ \\  
          \cmidrule{2-8}
          & Ours w/o TVD+TFS                             & $80.34_{\pm 6.75 (p < 0.001)}$   & & $5.947_{\pm 3.889 (p < 0.001)}$ & $3.777_{\pm 3.200 \, (p < 0.001)}$ & $2.170_{\pm 1.936\,  (p < 0.001)}$ &  $0.802_{\pm 0.990\,  (p < 0.001)}$\\ % & &  0 \\ & 38.118 & 
          & Ours w/o TVD                                 & $80.51_{\pm 6.66 (p < 0.001)}$   & & $2.957_{\pm2.396 (p < 0.001)}$  & $1.667_{\pm 1.828 \, (p < 0.001)}$ & $1.290_{\pm 1.278\,  (p = 0.004)}$ &    $0.629_{\pm 0.578\,  (p < 0.001)}$\\ % & &  0 \\& \textbf{29.360} &
          & Ours                                         & $\mathbf{81.02}_{\pm 6.30}$          & & $\mathbf{2.330}_{\pm 1.742}$ & $\mathbf{1.057}_{\pm 1.312}$ & $\mathbf{1.273}_{\pm 1.174}$ &   $0.629_{\pm 0.579}$\\ % & &  0 \\ & 29.828 &
    
    \midrule
    \multirow{9}[2]{*}{\rotatebox{90}{FIVES (2D)}} 
          & cl-Dice loss~\cite{shit2021cldice}           & $80.93_{\pm 16.31 \,  (p < 0.001)}$ & & $47.966_{\pm 23.010 \,  (p < 0.001)}$ & $42.529_{\pm 21.807 \, (p < 0.001)}$ & $5.438_{\pm 6.176\,  (p < 0.001)}$ &   $7.011_{\pm 18.422\,  (p < 0.001)}$ \\ %& &  0 \\ & 192.198 & 
          & Boundary loss~\cite{kervadec2019boundary}    & $81.21_{\pm 16.80 \,  (p < 0.001)}$ & & $44.921_{\pm 24.321 \,  (p < 0.001)}$ & $39.678_{\pm 22.819 \, (p < 0.001)}$ & $5.244_{\pm 6.162\,  (p = 0.001)}$ &   $6.322_{\pm 17.955\,  (p < 0.001)}$ \\ %& &  0 \\ & 180.633 & 
          & PH loss~\cite{hu2019topology}                & $81.15_{\pm 16.88 \,  (p < 0.001)}$ & & $39.345_{\pm 21.500 \,  (p < 0.001)}$ & $34.046_{\pm 20.063 \, (p < 0.001)}$ & $5.299_{\pm6.167\,  (p = 0.017)}$ &  $6.376_{\pm 18.164\,  (p < 0.001)}$ \\ %& &  0 \\ & 175.843 & 
          & Warp loss~\cite{hu2022structure}             & $81.31_{\pm 16.77 \,  (p < 0.001)}$ & & $56.395_{\pm 33.951 \,  (p < 0.001)}$ & $50.474_{\pm 30.113 \, (p < 0.001)}$ & $5.921_{\pm 7.074\,  (p < 0.001)}$ &   $6.158_{\pm 17.898\,  (p < 0.001)}$ \\ %& &  0 \\ & 181.972 & 

        % & Swin Unet~\cite{cao2022swin} & $80.32_{\pm 12.41 (p < 0.001)}$ & & $29.85_{\pm 14.41 (p < 0.001)}$ & $24.549_{\pm 12.846}$ & $5.301_{\pm 6.145}$ &  & $\mathbf{4.879}_{\pm 18.071}$ \\  
        & Swin-Unet~\cite{cao2022swin} & $79.54_{\pm 12.88 (p < 0.001)}$ & & $40.124_{\pm 17.453 (p < 0.001)}$ & $34.794_{\pm 15.467 \, (p < 0.001)}$ & $5.330_{\pm 6.506\,  (p < 0.001)}$ &   $6.148_{\pm 19.665\,  (p < 0.001)}$ \\   
         & MedSAM~\cite{ma2024segment} & $80.53_{\pm 14.13 \, (p < 0.001)}$ & & $33.808_{\pm 13.623 \, (p < 0.001)}$ & $28.793_{\pm 12.245 \, (p < 0.001)}$ & $5.015_{\pm 5.901\,  (p < 0.001)}$ &   $6.773_{\pm 25.092\,  (p < 0.001)}$ \\  
          \cmidrule{2-8}
          & Ours w/o TVD+TFS                             & $81.45_{\pm 16.57 \,  (p < 0.001)}$ & & $54.273_{\pm 31.352 \,  (p < 0.001)}$ & $48.733_{\pm 28.544 \, (p < 0.001)}$ & $5.540_{\pm6.526\,  (p < 0.001)}$ &   $6.193_{\pm 17.489\,  (p < 0.001)}$\\ % & &  0 \\ & 183.951 & 
          & Ours w/o TVD                                 & $83.36_{\pm 16.80 \,  (p < 0.001)}$ & & $28.440_{\pm 16.649 \,  (p < 0.001)}$ & $23.402_{\pm 14.872 \, (p < 0.001)}$ & $\mathbf{5.037}_{\pm 6.032\,  (p < 0.001)}$ &   $5.165_{\pm 16.875\,  (p < 0.001)}$ \\ % & \textbf{164.179} & 
          & Ours                                         & $\mathbf{84.07}_{\pm 15.53}$        & & $\mathbf{26.562}_{\pm15.255}$ & $\mathbf{21.399}_{\pm 13.609}$ & $5.164_{\pm 6.183}$ &   $\mathbf{5.039}_{\pm 16.080}$ \\ %& &  & 166.618 & 
    
    \midrule
    
    \multirow{9}[2]{*}{\rotatebox{90}{dHCP (2D)}} 
          & cl-Dice loss~\cite{shit2021cldice}           & $88.26_{\pm 5.94 \,  (p < 0.001)}$ & & $2.483_{\pm2.573 \,  (p < 0.001)}$ & $1.012_{\pm 1.389 \, (p < 0.001)}$ & $1.471_{\pm 1.914\,  (p < 0.001)}$ &   $0.230_{\pm 1.049\,  (p < 0.001)}$ \\ %& & 0.022 \\ & 6.412 & 
          & Boundary loss~\cite{kervadec2019boundary}    & $88.52_{\pm 5.64 \,  (p < 0.001)}$ & & $2.470 _{\pm2.574 \,  (p < 0.001)}$ & $0.999_{\pm 1.353 \, (p < 0.001)}$ & $1.471_{\pm 1.940\,  (p < 0.001)}$ &   $\mathbf{0.210}_{\pm 0.831\,  (p < 0.001)}$\\ % &&  0.024\\ & 6.413 & 
          & PH loss~\cite{hu2019topology}                & $88.17_{\pm 5.89 \,  (p < 0.001)}$ && $ 2.546 _{\pm2.676 \,  (p < 0.001)}$ & $1.075_{\pm 1.506 \, (p < 0.001)}$ & $1.471_{\pm 1.936\,  (p < 0.001)}$ &    $0.232_{\pm 1.134\,  (p < 0.001)}$\\ % &&  3.300 \\ & 6.466 &
          & Warp loss~\cite{hu2022structure}             & $88.11_{\pm 5.99 \,  (p < 0.001)}$ & & $2.495_{\pm 2.550 \,  (p < 0.001)}$ & $1.047_{\pm 1.418 \, (p < 0.001)}$ & $1.448_{\pm 1.870\,  (p < 0.001)}$ &   $0.230_{\pm 0.927\,  (p < 0.001)}$\\ % & & 0.339 \\ & 6.472 & 

         & Swin-Unet~\cite{cao2022swin} & $87.26_{\pm 6.43 (p < 0.001)}$ & & $3.852_{\pm 3.693 (p < 0.001)}$ & $2.280_{\pm 2.783 \, (p < 0.001)}$ & $1.572_{\pm 1.965\,  (p < 0.001)}$ &   $0.315_{\pm 1.180\,  (p < 0.001)}$ \\  
        & MedSAM~\cite{ma2024segment} & $88.43_{\pm 5.66 \, (p = 0.001)}$ & & $2.221_{\pm 2.355 \, (p < 0.001)}$ & $0.879_{\pm 1.293 \, (p < 0.001)}$ & $1.342_{\pm 1.750\,  (p < 0.001)}$ &   $\mathbf{0.210}_{\pm 0.858\,  (p < 0.001)}$ \\  
        
          \cmidrule{2-8}
          & Ours w/o TVD+TFS                             & $88.14_{\pm 5.87  \, (p < 0.001)}$ && $2.542_{\pm2.619  \, (p < 0.001)}$  & $1.091_{\pm 1.500 \, (p < 0.001)}$ & $1.451_{\pm 1.905\,  (p < 0.001)}$ &   $0.227_{\pm 0.834\,  (p < 0.001)}$ \\ %& & \textbf{0.021} \\ & 6.504 & 
          & Ours w/o TVD                                 & $88.45_{\pm 6.14  \, (p < 0.001)}$ &&  $2.220_{\pm 2.368  \, (p < 0.001)}$ & $0.824_{\pm 1.140 \, (p < 0.001)}$ & $1.395_{\pm 1.892\,  (p < 0.001)}$ &    $0.228_{\pm 0.922\,  (p < 0.001)}$ \\ %& & 0.042\\ & 6.348 &
          & Ours                                         & $\mathbf{88.56}_{\pm 6.26}$        &&  $\mathbf{2.032}_{\pm 2.182}$& $\mathbf{0.737}_{\pm 1.050}$ & $\mathbf{1.295}_{\pm 1.778}$ &   $0.218_{\pm 0.845}$ \\ %&  & 0.109\\  & \textbf{6.059} & 

    \midrule
    \multirow{9}[2]{*}{\rotatebox{90}{TopCoW (3D)}} 
          & cl-Dice loss~\cite{shit2021cldice}           & $94.48_{\pm 2.74 \, (p = 0.012)}$ & & $2.500_{\pm1.544 \, (p = 0.028)}$ & $1.433_{\pm 1.055 \, (p = 0.018)}$ & $1.067_{\pm 1.436\,  (p < 0.001)}$ &  $0.113_{\pm 0.110\,  (p < 0.001)}$ \\ %& &  0 \\  & 15.555 & 
          & Boundary loss~\cite{kervadec2019boundary}    & $94.55_{\pm 2.62 \, (p = 0.016)}$ & & $2.567_{\pm1.564 \, (p = 0.009)}$ & $1.600_{\pm 1.200 \, (p = 0.002)}$ & $\mathbf{0.967}_{\pm 1.402\,  (p < 0.001)}$ &   $0.110_{\pm 0.005\,  (p < 0.001)}$\\ % & &  0 \\ & 15.490 & 
          & PH loss~\cite{hu2019topology}                & $94.62_{\pm 2.41 \, (p = 0.454)}$ & & $2.600_{\pm1.705\,  (p = 0.018)}$ & $1.600_{\pm 1.405 \, (p = 0.001)}$ & $1.000_{\pm 1.461\,  (p = 0.005)}$ &    $\mathbf{0.099}_{\pm 0.079\,  (p=0.045)}$\\ % & &  0 \\ & \textbf{14.325} &
          & Warp loss~\cite{hu2022structure}             & $93.99_{\pm 3.23 \, (p < 0.001)}$ & & $2.900_{\pm2.071 \, (p < 0.001)}$ & $1.867_{\pm 1.454 \, (p < 0.001)}$ & $1.033_{\pm 1.494\,  (p = 0.529)}$ &   $0.130_{\pm 0.141\,  (p < 0.001)}$\\ % & &  0 \\  & 16.419 &

            & Swin-Unet~\cite{cao2022swin} & $92.62_{\pm 2.96 \, (p < 0.001)}$ & & $6.667_{\pm 3.458\,  (p < 0.001)}$ & $5.133_{\pm 2.837 \, (p < 0.001)}$ & $1.533_{\pm 1.454\,  (p < 0.001)}$ &   $0.181_{\pm 0.130\,  (p < 0.001)}$ \\  
            % & MedSAM~\cite{ma2024segment} & $93.94_{\pm 2.46 \, (p < 0.001)}$ & & $3.067_{\pm 2.097 \, (p < 0.001)}$ & $1.933_{\pm 1.153}$ & $1.133_{\pm 1.707}$ &  & $\mathbf{0.092}_{\pm 0.056}$ \\  
            % & MedSAM~\cite{ma2024segment} & $91.14_{\pm 4.20 \, (p < 3.299\times10^{-9})}$ & & $2.93_{\pm 2.82 \, (p < 0.0174)}$ & $1.90_{\pm 1.64}$ & $1.03_{\pm 1.64}$ &  & $\mathbf{0.113}_{\pm 0.062}$ \\  
            & MedSAM~\cite{ma2024segment} & $91.14_{\pm 4.20 \, (p < 0.001)}$ & & $2.933_{\pm 2.816 \, (p = 0.017)}$ & $1.900_{\pm 1.640\,  (p = 0.011)}$ & $1.033_{\pm 1.643\,  (p < 0.001)}$ &   $0.113_{\pm 0.062\,  (p < 0.001)}$ \\

          \cmidrule{2-8}
          & Ours w/o TVD+TFS                             & $94.28_{\pm3.11 \, (p = 0.016)}$ & & $2.800_{\pm2.072 \, (p < 0.001)}$ & $1.767_{\pm 1.453 \, (p < 0.001)}$ & $1.033_{\pm 1.779\,  (p < 0.001)}$ &   $0.106_{\pm 0.073\,  (p < 0.001)}$ \\ %& &  0 \\  & 15.987 &
          & Ours w/o TVD                                 & $94.61_{\pm2.54\,  (p <  0.001)}$ & & $2.700_{\pm1.828\,  (p < 0.001)}$ & $1.567_{\pm 1.283 \, (p < 0.001)}$ & $1.133_{\pm 1.707\,  (p < 0.001)}$ &   $0.106_{\pm 0.099\,  (p < 0.001)}$ \\%& &  %0 \\ & 14.882 & 
          & Ours                                         & $\mathbf{94.63}_{\pm2.52}$ & & $\mathbf{2.267}_{\pm1.731}$ & $\mathbf{1.300}_{\pm 1.130}$ & $\mathbf{0.967}_{\pm 1.683}$ &  $0.103_{\pm 0.094}$ \\%& &  %0 \\  & 14.683 &
    \bottomrule
    \end{tabularx}%
    }
  \label{tab:main}%
\end{table*}%

% figure 8
\begin{figure*}
% \vspace{-0.5cm}
\centering
    \centering
    \includegraphics[width=1.0\linewidth]{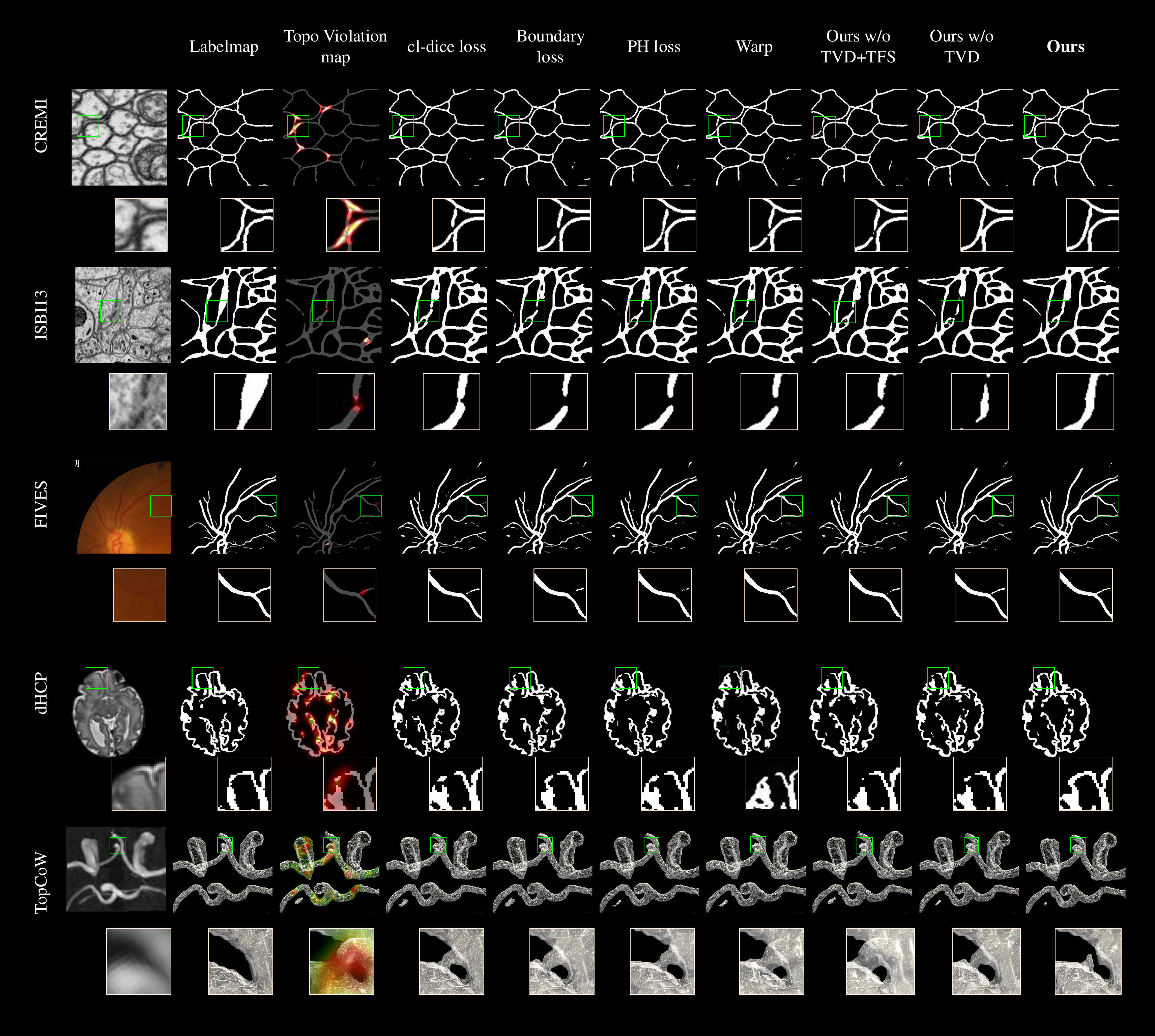}
    \caption{Generated topology violation maps and qualitative comparison of segmentation results. We compare our segmentation results with four baseline methods (cl-dice loss, Boundary loss, PH loss, and Warp loss) and two ablations in the last three columns. Our method can correct disconnected structures better than existing methods. }
\label{fig:exp}
% \vspace{-0.5cm}
\end{figure*}

% figure 9
\begin{figure*}[h]
\centering
\includegraphics[width=0.49\textwidth,trim={0 0cm 0 0},clip]{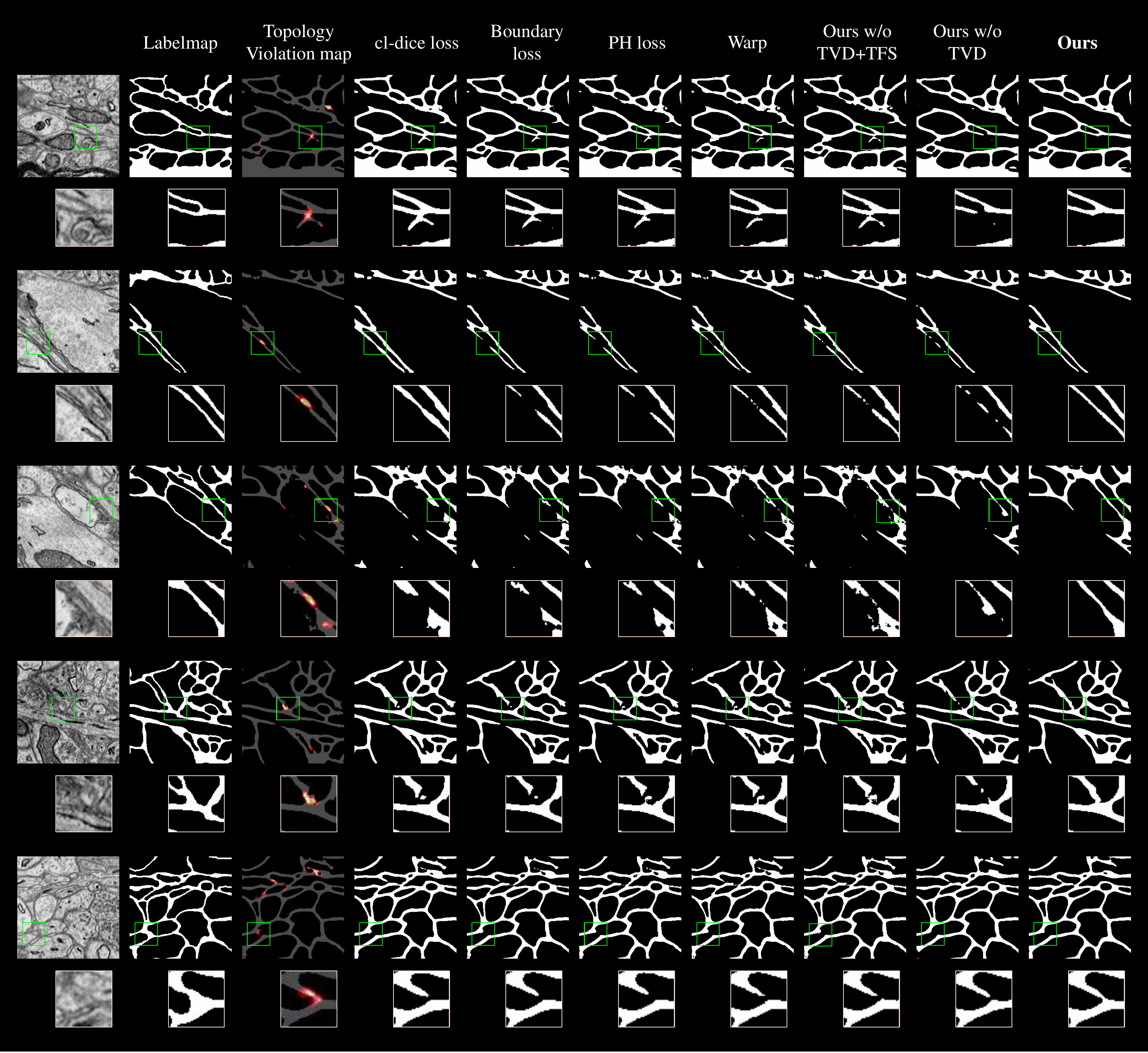}
\includegraphics[width=0.49\textwidth,trim={0 0cm 0 0},clip]{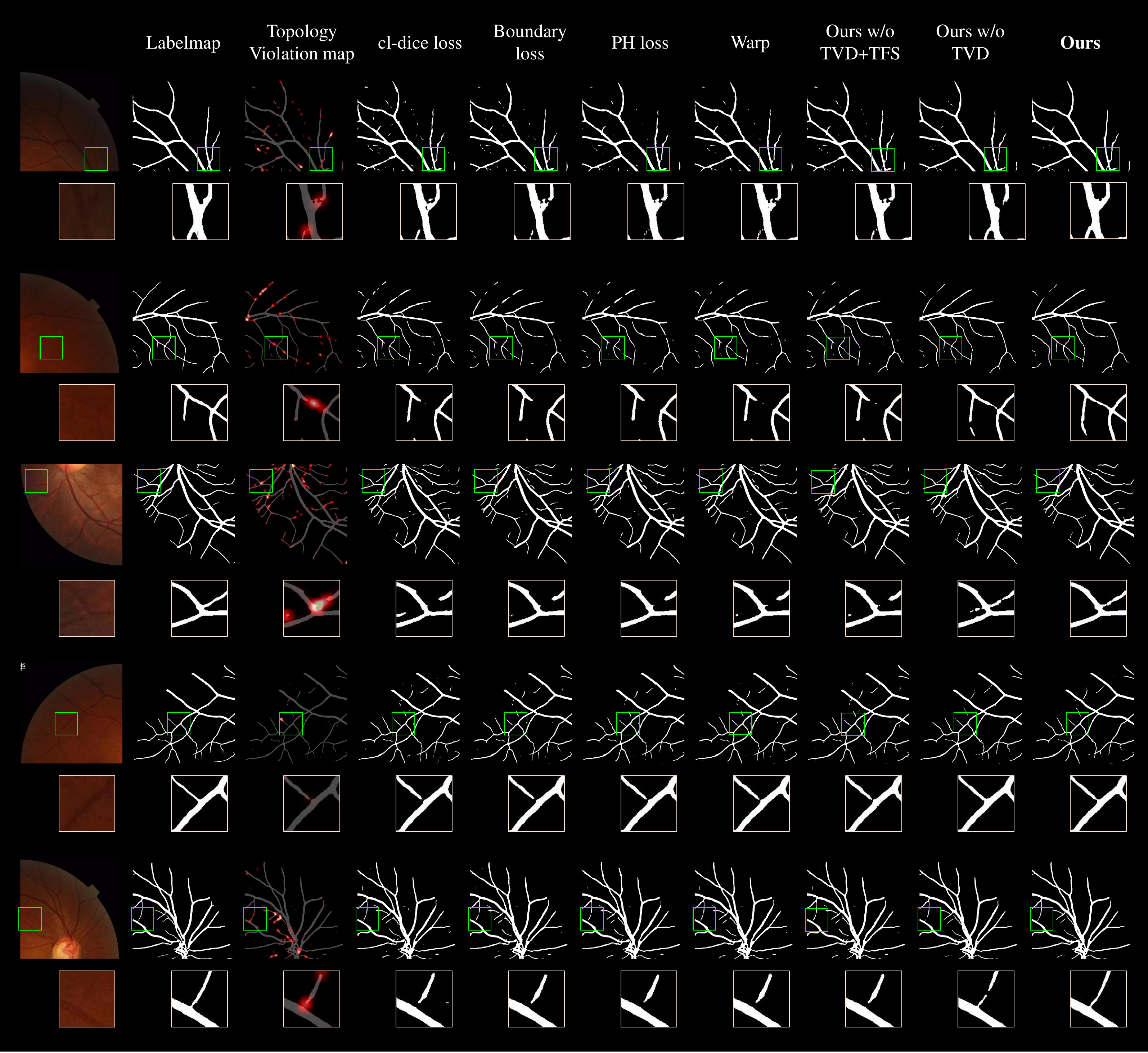}
\caption{Generated topology violation maps and the segmentation examples. Left: ISBI13 dataset, Right: FIVES dataset.} 
\label{fig:exp2}
\end{figure*}

\subsection{Datasets} 
We conducted experiments on the following datasets:
\begin{enumerate}

\item Circuit Reconstruction from Electron Microscopy Images (CREMI)~\cite{funke2018large} is a dataset with electron microscopy images for neuron boundary segmentation. It consists of 3 subsets, each of which consists of 125 $1250\times1250$ gray-scale images and corresponding label maps. We randomly selected 90 samples from each subset (270 samples in total) as the training set, 10 samples from each subset (30 samples in total) as the validation set, and use the remaining 25 samples (75 samples in total) as test set. We further divided each $1250\times1250$ image into 25 $256\times256$ patches with an overlap of 8 pixels, in order to fit the GPU memory and enlarge the size of the training set. Thus, training, validation, and test sets consist of 6,750, 750 and 1,875 samples, respectively.

\item ISBI13~\cite{arganda2013snemi} consists of 100 electron microscopy images with neurites segmentation ground-truth. The resolution of each slice is $1024\times1024$. To enlarge the dataset, we crop each slice into 16 $256\times256$ patches, and split it into 1200, 100, and 300 for training, validation and test, respectively.

\item Fundus Image Vessel Segmentation (FIVES) dataset~\cite{jin2022fives} is a dataset of fundus images for retinal vessel segmentation. The dataset consists of 800 samples, and the origin resolution for each sample is $2048\times2048$. To enlarge the dataset and keep the basic topological structure of vessels, we crop each slice into 4 $1024\times1024$ patches. In total we have 2000, 400 and 800 slices for training, validation and test, respectively.

\item The Developing Human Connectome Project (dHCP$\footnote{\scriptsize \href{www.developingconnectome.org}{www.developingconnectome.org}}$) dataset has 242 fetal brain T2 Magnetic Resonance Imaging (MRI) scans with gestational ages ranging from $20.6$ to $38.2$ weeks. All MR images were motion corrected and reconstructed to $0.8$ mm isotropic resolution for the fetal head region of interest (ROI) \cite{kuklisova2012reconstruction,kainz2015fast}. The images are affinely aligned to the MNI-152 coordinate space and clipped to the size of $144 \times 192\times 192$. We randomly split the data into 145 samples for training, 73 for testing, and 24 for validation. 
We conduct cortical gray matter segmentation on this dataset that the label is first generated by DrawEM method~\cite{makropoulos2018developing} and then refined manually to improve the segmentation accuracy.

\item Topology-Aware Anatomical Segmentation of the Circle of Willis (TopCoW) dataset is the first 3D dataset with voxel-level annotations for the vessels of the Circle of Willis~\cite{yang2023benchmarking}. It contains magnetic resonance angiography (MRA) data with a magnetic field strength of 3 Tesla or 1.5 Tesla. 
We randomly split the 90 publicly available labeled volumes into 55 for training, 5 for validation, and 30 for testing. The training and evaluation are done on the ROI regions provided.
\end{enumerate}

% We conduct experiments on two datasets: CREMI for neuron boundary segmentation~\cite{funke2018large} and  
These datasets are used to evaluate different topology challenges, where neuron boundaries in CREMI~\cite{funke2018large} and ISBI13~\cite{arganda2013snemi} have a random and diverse number of holes $\beta_1$ and connected components $\beta_0$, while the cortex in the dHCP dataset is expected to have fixed $\beta_1$ and $\beta_0$, and the vessel structure in FIVES~\cite{jin2022fives} and TopCoW~\cite{yang2023benchmarking} should remain connected tubular structures during segmentation.

\subsection{Implementation} 
% The hyper-parameters in Eq.~\ref{eq:3} and Eq.~\ref{eq:4} are empirically chosen as ${\rm {\Delta}}=32$ for the CREMI dataset, ${\rm {\Delta}}=8$ for the dHCP dataset, and $t=0.6$ for both datasets. We choose ${\rm {\Delta}}$ to be higher for CREMI than for dHCP because: (1) the resolution of CREMI is higher and (2) the topology of the fetal cortex may change in smaller regions. 
% We choose a U-Net as the backbone for all methods for comparison. However, due to the modular nature of our method it can also be incorporated into other segmentation frameworks. 
The architecture of the TFS network uses skip connection like a U-Net~\cite{ronneberger2015u}, with 3 blocks of the encoder layers and 3 blocks of the decoder layers. Each encoder block applies a downsampling operation, progressively reducing the spatial resolution to $1/8$ of the original input size, with the bottleneck feature containing 64 channels. ReLU activation is used in all convolutional layers except the final layer, which applies a sigmoid activation for segmentation. The TFS network is trained by the cross-entropy loss.

The training process of our pipeline is two steps, i.e., the segmentation network is first trained to obtain coarse segmentations, then the TFS network.
Note that the parameters in TVD are fixed. We apply the Adam optimizer~\cite{kingma2014adam} with a weight decay of $1\times e^{-4}$ to train the TFS network. The initial learning rate is set to $1\times e^{-4}$ and is reduced by a factor of 2 every 50 epochs. The model is trained for 200 epochs until converged.

We choose U-Net as the segmentation backbone for our method and for cl-Dice loss~\cite{shit2021cldice}, boundary loss~\cite{kervadec2019boundary}, warp loss~\cite{hu2022structure} and PH loss~\cite{hu2019topology} for comparison. 
However, due to the modular nature of our method it can also be incorporated into other segmentation frameworks, as discussed in Section IV-F Ablation study.

We use PyTorch 1.13.1 and calculate the Betti number with the GUDHI package \cite{maria2014gudhi}. The training time is evaluated on an NVIDIA RTX $3080$ GPU with a batch size of 20. 

\subsection{Evaluation metrics} 
Segmentation performance is evaluated by Dice score and averaged surface distance (ASD), and the performance of topology is evaluated by Betti errors, which is defined as: %the difference $e_i$ between prediction Betti number $\beta^{\mathrm{pred}}_{i}$ and GT Betti number $\beta^{\mathrm{gt}}_{i}$: 
$e_i = \mid \beta^{\mathrm{pred}}_{i} - \beta^{\mathrm{gt}}_{i}\mid$, where $i\in \{0,1\} $ indicates the dimension. %Note that for 2D images, $\beta_{i}=0$ when $i\ge2$. 
We also report the mean Betti error as $e = e_0 + e_1$.

\subsection{Quantitative evaluation}
We compare the segmentation performance of our approach against six baseline methods. Among them, four are designed to preserve shape and topology: clDice loss~\cite{shit2021cldice}, Boundary loss~\cite{kervadec2019boundary}, Warp loss~\cite{hu2022structure}, and PH loss~\cite{hu2019topology}. Additionally, we compare our method with two segmentation pipelines: Swin-Unet~\cite{cao2022swin} and MedSAM~\cite{ma2024segment}.
For pixel-wise accuracy, our method achieves the best Dice score on all five datasets, as demonstrated in Tab.~\ref{tab:main}. Despite the comparable ASD scores, we achieve significant improvements in terms of Betti errors. 
Compared to all baseline methods, our method achieves better topological accuracy in terms of Betti error. For example, the average Betti error of our method is $4.941$ on the CREMI dataset, representing an improvement of $3.853$ ($43.8\%$) over MedSAM and 2.186 ($30.7\%$) over Swin-Unet in mean Betti error. We observe similar improvements for the ISBI13, FIVES, dHCP, and TopCoW datasets.
We conducted Wilcoxon signed-rank tests~\cite{wilcoxon1992individual} to compare our method against other baselines on all five datasets. As shown in Tab.~\ref{tab:main}, our method is statistically better than the baseline methods with in terms of Betti error with all p-values below 0.05.

The averaged training times of our method is 0.0134 s/item on CREMI dataset, with 0.0168 s/item, 0.0151 s/item, 0.1033 s/item, and 0.9748 s/item for cl-Dice loss~\cite{shit2021cldice}, boundary loss~\cite{kervadec2019boundary}, warp loss~\cite{hu2022structure} and PH loss~\cite{hu2019topology}, respectively. Compared to the PH-based loss \cite{hu2019topology}, which is a similarly well-grounded concept in algebraic topology, our method is 72.7 times faster than the PH-based loss \cite{hu2019topology}.
 % apart from PH loss~\cite{hu2019topology}, our method is the only approach designed directly based on topology theory. However, the computation time for PH loss is 8.772 s/batch on the CREMI dataset, whereas our method is 46.4 times faster at 0.189 s/batch.

To assess the effect of end-to-end training, we conducted additional experiments on the ISBI13~\cite{arganda2013snemi} dataset, where the segmentation and TFS networks were trained jointly by updating the parameters of the segmentation network and then updating the parameters of the TFS network in each iteration. As shown in Tab.~\ref{tab:end-to-end}, this approach consistently reduced Betti error while maintaining the Dice similarity of TFS-refined segmentations. The improvement can be attributed to the TFS network encountering a wider range of topological errors during early training stages, enhancing its ability to refine segmentation outputs. These results suggest that end-to-end training can be further beneficial.

\begin{figure*}[t]
\centering
\includegraphics[width=0.31\textwidth,trim={0 0cm 0 0},clip]{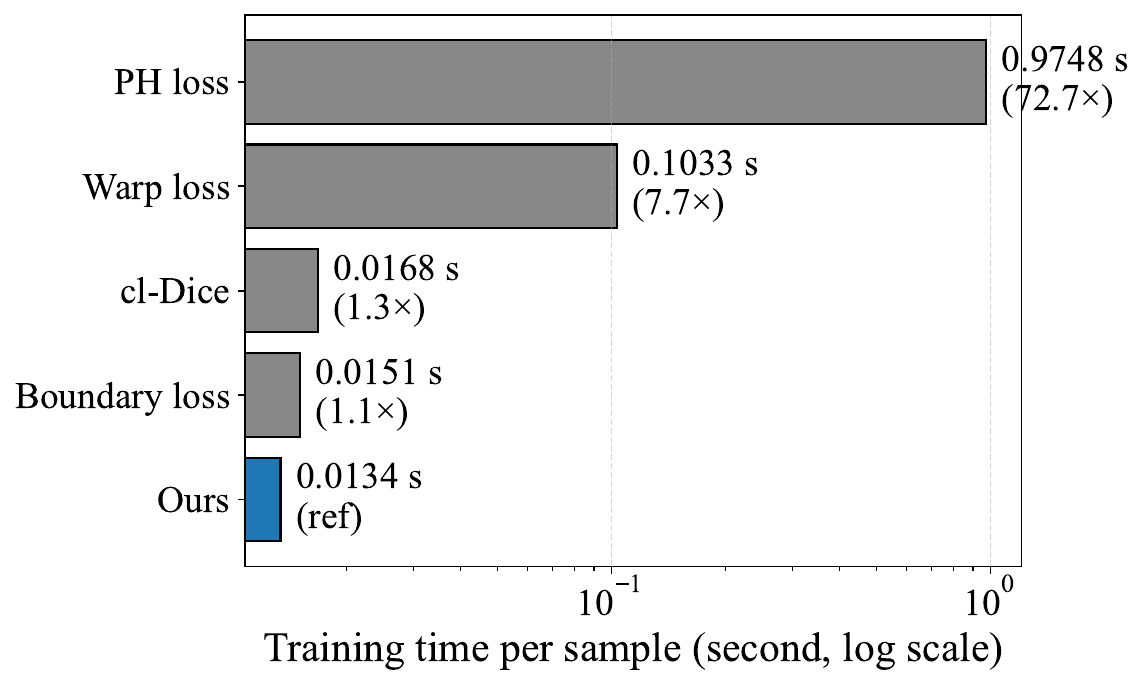}
\includegraphics[width=0.31\textwidth,trim={0 0cm 0 0},clip]{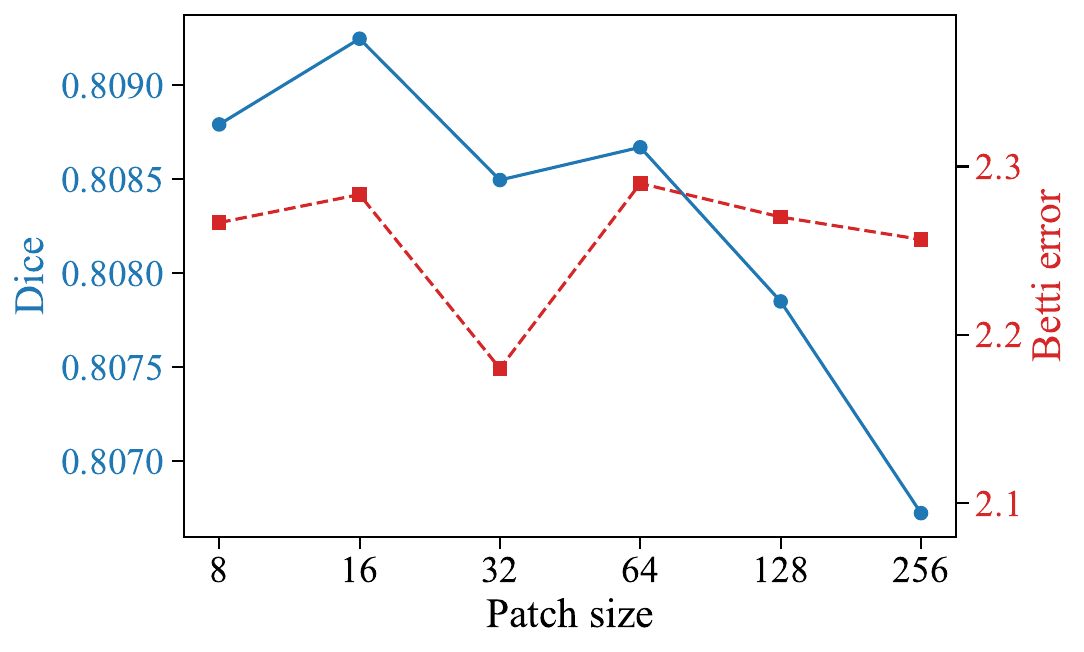}
\includegraphics[width=0.31\textwidth,trim={0 0cm 0 0},clip]{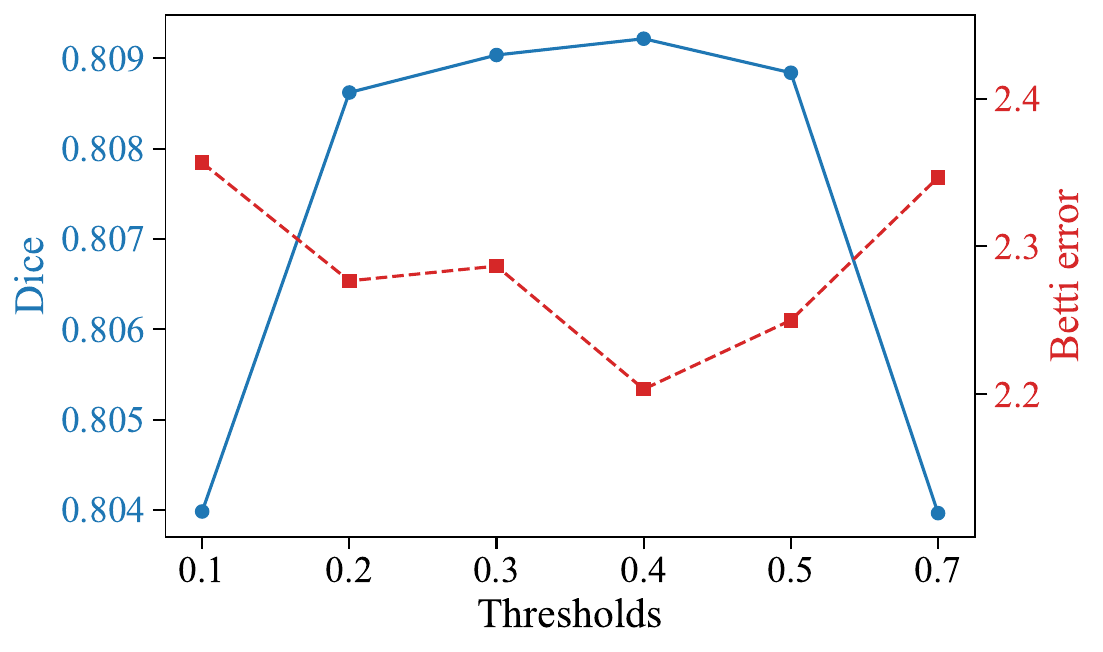}
\caption{Left: comparison of training time per sample (in seconds) on the CREMI dataset, shown on a logarithmic scale. Middle and right: ablation study of different patch sizes and decision thresholds in the TVD block on the ISBI13 dataset.
}
\label{fig:ablation}
\end{figure*}

% table 2
\begin{table*}[t!]
\centering
% \scriptsize
\caption{Comparison of training strategies: end-to-end training vs. sequential training for pre-segmentation and post-topology correction networks}
\label{tab:training_comparison}
\begin{tabularx}{\textwidth}{cp{2cm}XcXXXcX}
\toprule
 & Training type & Dice $\uparrow$ & & $e \downarrow$ & $e_0 \downarrow$ & $e_1 \downarrow$ & & ASD $\downarrow$ \\
\midrule
\multirow{2}{*}{Pre-segmentation} 
& Two-stage  & $\mathbf{80.34}_{\pm 6.75}$ & & $5.947_{\pm 3.889}$  & $3.777_{\pm 3.200}$  & $2.170_{\pm 1.936}$  & &$\mathbf{0.802}_{\pm 0.990}$ \\
& End-to-end  & $79.00_{\pm 7.18}$  && $\mathbf{4.750}_{\pm 2.904}$  & $\mathbf{3.010}_{\pm 2.413}$  & $\mathbf{1.740}_{\pm 1.602}$  & &$0.910_{\pm 1.194}$ \\

\midrule
\multirow{2}{*}{Post-TFS refined} 
& Two-stage   & $\mathbf{81.01}_{\pm 6.30}$ & & $2.330_{\pm 1.742}$  & $1.057_{\pm 1.312}$  & $1.273_{\pm 1.174}$  & &$0.629_{\pm 0.579}$ \\
& End-to-end   & $80.98_{\pm 6.41}$  & & $\mathbf{2.147}_{\pm 1.492}$  & $\mathbf{0.943}_{\pm 1.010}$  & $\mathbf{1.203}_{\pm 1.150}$  && $\mathbf{0.660}_{\pm 0.752}$ \\
\bottomrule
\end{tabularx}%
\label{tab:end-to-end}
\end{table*}

% table 3
\begin{table*}[t!]
% \scriptsize
    \centering
    \caption{Comparison of segmentation backbones: incorporating our TVD and TFS networks into UNet~\cite{ronneberger2015u} and nnU-Net~\cite{isensee2021nnu}.}
    \begin{tabularx}{\textwidth}{cp{3cm}XcXXXcX}
    \toprule
       Data & Method & Dice $\uparrow$ & &$e$ $\downarrow$& $e_0$ $\downarrow$ & $e_1$  $\downarrow$ &  & ASD $\downarrow$  \\ 
          \midrule

          \multirow{4}{*}{\rotatebox{90}{ISBI13}} 
          & U-Net & $80.34_{\pm 6.75}$ & & $5.947_{\pm 3.889}$ & $3.777_{\pm 3.200}$ & $2.170_{\pm 1.936}$ & & $0.802_{\pm 0.990}$\\ 
          & U-Net + Ours & $81.02_{\pm 6.30}$ & & $2.330_{\pm 1.742}$ & $1.057_{\pm 1.312}$ & $1.273_{\pm 1.174}$ &  & $0.629_{\pm 0.579}$\\ 
          & nnU-Net &  $84.02_{\pm 5.55}$ & & $1.307_{\pm 1.125}$ & $0.470_{\pm 0.665}$ &  $0.837_{\pm 0.900}$ & &  $0.366_{\pm 0.473}$ \\
          & nnU-Net + Ours &  $\mathbf{84.23}_{\pm 5.49}$ & & $\mathbf{1.243}_{\pm 1.091}$ & $\mathbf{0.457}_{\pm 0.684}$ &  $\mathbf{0.787}_{\pm 0.865}$ & &  $\mathbf{0.365}_{\pm 0.490}$ \\

        \midrule
          \multirow{4}{*}{\rotatebox{90}{TopCoW}} 
          & U-Net & $94.28_{\pm 3.11}$ & & $2.800_{\pm2.072}$ & $1.767_{\pm 1.453}$ & $1.033_{\pm 1.779}$ &  & $0.106_{\pm 0.073}$ \\ 
          & U-Net + Ours & $94.63_{\pm 2.52}$ & & $2.267_{\pm 1.731}$ & $1.300_{\pm 1.130}$ & $0.967_{\pm 1.683}$ &  & $0.103_{\pm 0.094}$ \\ 
          & nnU-Net &  $\mathbf{95.15}_{\pm 3.02}$ & & $1.700_{\pm 1.696}$ & $\mathbf{0.700}_{\pm 0.781}$ &  $1.000_{\pm 1.693}$ & &  $\mathbf{0.074}_{\pm 0.072}$ \\
          & nnU-Net + Ours &  $94.81_{\pm 2.66}$ & & $\mathbf{1.433}_{\pm 1.230}$ & $0.733_{\pm 0.727}$ &  $\mathbf{0.700}_{\pm 1.159}$ & &  $\mathbf{0.074}_{\pm 0.066}$ \\
    \bottomrule
    \end{tabularx}%
    \label{tab:nnUNet}
\end{table*}

\subsection{Qualitative evaluation}
 As highlighted in Fig.~\ref{fig:exp} and Fig.~\ref{fig:exp2}, our method can effectively eliminate the topological errors in both 2D and 3D datasets, outperforming all the other methods. For instance, in the first row of the CREMI dataset, none of the baseline methods could segment the tiny neuron structure when the boundary of the cells is blurry. Similarly, in the dHCP dataset, all the baseline methods fail to segment the fetal cortex as a closed surface, whereas our method can successfully resolve this issue. Also in FIVES dataset, our method can capture the tiny vessel structure as an connected component while all other methods fails. In 3D TopoCoW dataset, 
 Second, we show the topology violation maps from the TVD block in the third column in Fig.~\ref{fig:exp}, which indicate the topology error regions between the GT (second column) and the prediction from our first segmentation network (eighth column). For example, in the evaluation on TopCoW dataset, we observe that the topology violation map can highlight the hole structure with topological error, therefore driving our method to correct these errors. Similar results can be observed on the example of other datasets.

\subsection{Ablation study} 
We first evaluate the effectiveness of our TVD design. We train our pipeline without the TVD block. Instead, we use the difference map between the prediction and GT as a substitute for the topology violation map. Qualitative results are shown in Tab.~\ref{tab:main} between rows `Ours w/o TVD' and `Ours' across five datasets. Ours statistically outperforms the ablation setting of removing the TVD block in terms of Dice similarity and Betti errors, with p-value consistently lower than 0.001, demonstrating the effectiveness of our TVD design. We also summarize the qualitative results in Fig.~\ref{fig:exp} and Fig.~\ref{fig:exp2}. %For example, %as the highlight in the first sample of the CREMI dataset, 
Feature synthesis with the difference map can correct some of the false negative errors, however, structures tend to remain incorrect for all datasets. In contrast, our approach successfully corrects these topological errors. %Similar results are observed in the rest samples. 
% Quantitative results are also provided in Table~\ref{tab:main}. Our method with TVD outperforms the ablation study in terms of all metrics for both datasets.
Secondly, we further remove the TFS network. Quantitative results are provided in Tab.~\ref{tab:main}. The topology performance without TVD+TFS is significantly inferior to our method in terms of Betti errors, which illustrates the effectiveness of our design.

Additionally, we conducted an ablation study to assess the impact of different decision thresholds on segmentation performance using the ISBI13~\cite{arganda2013snemi} dataset. As shown in Fig.~\ref{fig:ablation},  Dice similarity first increases and then decreases as the threshold increases, while Betti error follows the opposite trend. At a high threshold $t = 0.7$, only a small fraction of pixels are masked and further corrected by the TVD block. Conversely, a low threshold ($t=0.1$) results in excessive masking, leading to insufficient input information for TVD to refine errors.  The best performance is observed when $t$ is between 0.2 and 0.5. Thus, we sample $t$ from this range during training to enhance the diversity of synthetic data.

To assess the impact of patch size on topology refinement, we conducted an ablation study using patch sizes of 8, 16, 32, 64, 128, and 256 on the ISBI13 dataset. As shown in Fig.~\ref{fig:ablation}, Betti error initially decreases and then increases with larger patches, while Dice similarity declines monotonically for patch sizes above 64. This trend reflects a trade-off in local topological representation. Larger patches may cause topological errors in different regions to cancel out, while overly small patches introduce local Euler characteristic variations due to boundary mismatches between predictions and ground truth, despite the global topology remaining unchanged. We thereby selected a patch size of 32 as the optimal hyperparameter in our experiments.

To evaluate the generalizability of our method, we conduct experiments with an alternative segmentation backbone by replacing UNet~\cite{ronneberger2015u} with nnU-Net~\cite{isensee2021nnu}. We trained the TFS network using the predictions from nnU-Net. As shown in Tab.~\ref{tab:nnUNet}, our method consistently improves the segmentation performance in terms of Betti error on both ISBI13~\cite{arganda2013snemi} and TopCoW~\cite{yang2023benchmarking} datasets, while maintaining Dice similarity. These results demonstrates the robustness of our design regardless of the segmentation backbone.

\subsection{Discussion} 
% fast
% no need to retrai

We illustrate the effectiveness of our approach across five distinct datasets featuring varied topological structures. Our results uniformly show that our method not only addresses topological inaccuracies but also surpasses all comparative baseline methods in both quantitative and qualitative assessments. Additionally, we use the numerical Betti error for topological evaluation and employ a topology violation map derived from $\chi$ to provide a visual representation of topological accuracy. This visual tool supports our TFS network in mitigating topological errors. An ablation study further substantiates the robustness of our method's design.

However, our approach retains a vulnerability to noise, as clearer image features and stronger image gradients yield better results. Typically, all topology-aware methods are also prone to errors from labelling noise; even minor mislabeling within an object can significantly disrupt topology. Strategies to mitigate this include label quality control or verification by multiple experts, but these measures are costly. Our method can be adapted to anticipate specific topological structures, such as being homeomorphic to a sphere, where the expected $\chi$ and Betti numbers are predefined.

We can extend our technique to address multi-class segmentation tasks by calculating $\chi$ for each class individually, similar to PH methods. While the computational time for both $\chi$ and PH increases linearly with more categories, our approach is more efficient per class and this efficiency is further amplified in multi-class segmentation.

Addressing topology characterization within non-linear yet Lipschitz continuous frameworks, like deep learning-based image segmentation, poses significant challenges. Counting functions are typically neither continuous nor differentiable. We overcome this by implementing fixed forward transformations with bit-quads and bit-octets to determine Betti numbers and $\chi$. 
Learning directly through backpropagation may be more elegant, but this results in non-local image-level error functions and relies on the computationally intensive continuous differentiation of filtration levels, \emph{e.g.}, in strategies based on PH. Thus, further exploration into differentiable and localised representations of $\chi$ could be a fruitful future research direction.

\iffalse
We demonstrate the effectiveness of our method on five datasets with different topological structures. The results consistently indicate that our method can correct topological errors and outperforms all the baseline methods both quantitatively and qualitatively. In addition to the numerical Betti error to evaluate the topological performance, the topology violation map computed from $\chi$ provides a visual impression of the topological performance, which further assists our TFS network to correct these topological errors. The ablation study further validates the design of our approach. 

A limitation of our method is its remaining sensitivity to noise. However, we observe better results for data with clearer structures and stronger image gradients. 
\fi

%Our efficient, easy deployed and 2D/3D compatible method

% This study sheds new light on the optimization and evaluation for medical image segmentation. We observe that most existing methods do not take topological constraints into account, therefore the predictions would be unconnected and unrealistic. As a computation-efficient block, our proposed method can be easily integrated into existing segmentation methods to constraints the topological structure. 

%In medical image segmentation tasks, when the Dice score has generally achieved its upper-bound for most of the models, how could we set our goal for further optimizing the model? This paper brings the idea of chasing for the correctness of topology. 

\section{Conclusion}
In this work, we propose a method aimed at improving and evaluating topology-aware medical image segmentation. We observe that most existing methods either do not consider topological information, or are limited by their high computational complexity. As a computationally efficient approach, our method can be easily integrated into existing segmentation methods to improve the topological structure. 

%they might be more suitable to datasets with clear topology structures.

We propose a novel $\chi$-based method to include topology constraints in the segmentation network. Different from PH-based approaches, our method has a distinct advantage in computational efficiency while providing improved performance. 

We also generate a topology violation map, which can aid  interpretability, and use this map directly in a post-processing feature synthesis network. %We believe this map is valuable and could be explored for other scenarios, such as serving as a spatial prior to regularize various loss functions. 
We believe this map is valuable and can be applied to other tasks. In image classification and disease diagnosis, topological structures could serve as discriminative features. Additionally, in generative tasks where topological priors are required, such as histopathology image generation~\cite{abousamra2023topology,xu2024topocellgen}, topology violation map can be used to assess the topological correctness of generated images and regularize loss functions accordingly.

Future work will focus on clinical translation and downstream evaluation for surface modelling and, \emph{e.g.}, flow or growth simulation applications. 

%TODO...
%In the future, we will consider to directly optimize the entire network to further reduce the EC error. 
%In the future, as the development of TVD is fully differentiable, it could also be possible to include this EC into a loss function as an explicit supervision.
%, however, the nature of EC which is the relation of $\beta_0$ and $\beta_1$ brings the limitation of using EC as a strong supervision signal as other ph-based loss function designs. 

%\noindent\textbf{Acknowledgements:} 
%This project is supported by Lee Family Scholarship from Imperial College London. HPC resources are provided by the Erlangen National High Performance Computing Center (NHR@FAU) of the Friedrich-Alexander-Universität Erlangen-Nürnberg (FAU) under the NHR project b143dc. NHR funding is provided by federal and Bavarian state authorities. NHR@FAU hardware is partially funded by the German Research Foundation (DFG) – 440719683. Support was also received by the ERC - project MIA-NORMAL 101083647 and DFG KA 5801/2-1, INST 90/1351-1.

%, more applications of this violation map is open to the readers, e.g., as a spatial prior to weight the cross-entropy loss. Moreover,  our method is only developed in 2d because of the Gray algorithm is designed in 2d, but we believe its not difficult to find a 3d kernel pattern to make this method fits in 3d. 
%\newpage
\bibliographystyle{IEEEtran}
\bibliography{mybib}

\end{document}